\documentclass[a4paper,11pt]{article}
\pdfoutput=1 
\usepackage{jheppub}	
\usepackage[utf8]{inputenc}
\usepackage[english]{babel}
\usepackage{makeidx}
\usepackage{amsfonts}
\usepackage{enumerate}
\usepackage{mathrsfs}
\usepackage{tensor}
\usepackage[autostyle]{csquotes}
\usepackage{subfig}
\newcommand{\ft}[2]{{\textstyle\frac{#1}{#2}}}

\newdimen\tableauside\tableauside=1.0ex
\newdimen\tableaurule\tableaurule=0.4pt
\newdimen\tableaustep
\def\phantomhrule#1{\hbox{\vbox to0pt{\hrule height\tableaurule
width#1\vss}}}
\def\phantomvrule#1{\vbox{\hbox to0pt{\vrule width\tableaurule
height#1\hss}}}
\def\sqr{\vbox{%
  \phantomhrule\tableaustep
\hbox{\phantomvrule\tableaustep\kern\tableaustep\phantomvrule\tableaustep}%
  \hbox{\vbox{\phantomhrule\tableauside}\kern-\tableaurule}}}
\def\squares#1{\hbox{\count0=#1\noindent\loop\sqr
  \advance\count0 by-1 \ifnum\count0>0\repeat}}
\def\tableau#1{\vcenter{\offinterlineskip
  \tableaustep=\tableauside\advance\tableaustep by-\tableaurule
  \kern\normallineskip\hbox
    {\kern\normallineskip\vbox
      {\gettableau#1 0 }%
     \kern\normallineskip\kern\tableaurule}%
  \kern\normallineskip\kern\tableaurule}}
\def\gettableau#1 {\ifnum#1=0\let\next=\null\else
  \squares{#1}\let\next=\gettableau\fi\next}
\def\be{\begin{equation}}
\def\ee{\end{equation}}
\def\bea{\begin{eqnarray}}
\def\eea{\end{eqnarray}}
\newcommand{\nn}{\nonumber}

\title{Chaos at the rim of black hole and fuzzball shadows
}
\author[a,b]{M. Bianchi,}
\author[a,b]{A. Grillo,}
\author[b]{J.F. Morales}
\affiliation[a]{Dipartimento  di  Fisica,  Universit\`a  di  Roma  ``Tor  Vergata'',
\\Via della Ricerca Scientifica, 00133 Roma, Italy }  
\affiliation[b]{I.N.F.N.  Sezione  di  Roma  ``Tor  Vergata'',
\\Via della Ricerca Scientifica, 00133 Roma, Italy}
\emailAdd{massimo.bianchi@roma2.infn.it}
\emailAdd{morales@roma2.infn.it}
\emailAdd{alfredo.grillo@roma2.infn.it}

\abstract{We study the scattering of massless probes in the vicinity of the {\it photon-sphere} of asymptotically AdS black holes and horizon-free microstate geometries (fuzzballs). We find that these exhibit a chaotic behaviour characterised by exponentially large deviations of nearby trajectories. We compute the Lyapunov exponent $\lambda$ governing the exponential growth in $d$ dimensions and show that it is bounded from above by $\lambda_b = \sqrt{d{-}3}/2b_{\rm min}$ where $b_{\rm min}$ is the minimal impact parameter under which a massless particle is swallowed by the black hole or gets trapped in the fuzzball for a very long time. Moreover we observe that   $\lambda$ is typically below the advocated bound on chaos $\lambda_H=2\pi \kappa_B T/\hbar$, that in turn characterises the radial fall into the horizon, but the bound is violated in a narrow window near extremality, where the photon-sphere coalesces with the horizon. Finally, we find that fuzzballs are characterised by  Lyapunov exponents  smaller than those of the corresponding BH's suggesting the possibility of discriminating the existence of micro-structures at horizon scales via the detection of ring-down modes with time scales $\lambda^{-1}$ longer than those expected for a BH of the given mass and spin.}


\preprint{PREPRINT}
 \clearpage

\begin{document}
\maketitle


 \section{Introduction}
 \label{Intro}
 
 A light ray travelling near a massive object gets deflected and delayed with respect to its propagation in flat space-time. If the massive object is a black hole (BH), the time delay and angular deflection, that are  small at large impact parameter, can get arbitrarily large at some critical value. For critical impact parameters even light gets trapped in (unstable) `circular'\footnote{We put `circular' in quotes, since closed geodesics at constant `radial' variable in a curved geometry are not necessarily circles.}  orbits around the BH. Below this critical threshold no signal can escape from the BH gravitational potential.
 
The limiting `circular' photon orbits form the BH {\it photon-sphere} that discriminates between the scattering and absorption regimes. It appears as the rim of the {\it black hole shadow} projected by the BH on a distant observer.  The size and shape of the photon-sphere depends on the mass, charge and angular  momentum of the BH and can help to uniquely identify it. For instance, for a Schwarzchild BH the photon-sphere is a sphere of radius 3/2 times that of the BH horizon\footnote{We observe that this radius is larger than the ``Buchdahl bound'' $R_B={9\over 8} r_H$, i.e. the outermost radius of stability for a fluid sphere made of `conventional' matter \cite{Buchdahl:1959zz}, so in principle even a star (more likely a cold neutron star than a hot star like our sun) can have a photon-sphere (not to be confused with the `photo-sphere').}. Photon orbits are also useful to explain the optical appearance of stars undergoing gravitational collapse and the response of the BH to perturbations, the so called quasi-normal modes, that can be interpreted as null particles trapped at the unstable
orbit and slowly leaking out \cite{Cardoso:2008bp}.

 Instability of the photon orbits signals the onset of chaos on the geodetic motion near the photon-sphere \cite{Cardoso:2008bp}. A measure of chaos or, better, of the high sensitivity to the initial conditions is given by the Lyapunov exponent $\lambda$,  that can be defined via the Poiss\`on bracket 
      \begin{equation}
\begin{aligned}
\{ \phi(0), \phi( t) \}_{\rm P.B}= \frac{\delta \phi(t)}
{ \delta P_\phi }
 \sim e^{\lambda\, t} \label{chaos}
\end{aligned}
\end{equation}
describing the exponential growth with time of the deflection along an angular direction $\phi$, for small variations of the corresponding conjugate momentum $P_\phi$.  In the case of spherically symmetric BH's, the Lyapunov exponent $\lambda$ can be related to the time decay of the basic QNM \cite{Cardoso:2008bp}. In an asymptotically AdS space, the chaotic behaviour manifests as a random angular dispersion of nearby trajectories re-emerging at the boundary after a long time $ t$, much longer than the time it takes for a free massless particle to travel from/to the boundary and back.  
  
  Our work is inspired by the results in \cite{Maldacena:2015waa}, where holography was used to diagnose chaos in quantum mechanical systems at finite temperature. The amount of chaos in the quantum theory was characterised by the exponential growth of  out-of-time-ordered correlators, 
involving commutators  that replace the Poiss\`on brackets in (\ref{chaos}) and encode the perturbation induced by an operator on a later measurement. The long time behaviour of the field theory correlator was related to the exponential growth $e^{\lambda_H t}$ of the center-of-mass energy of the high energy scattering process in the dual gravity theory \cite{Shenker:2013pqa, Shenker:2013yza}. The exponent $\lambda_H=2\pi \kappa_B T/\hbar$, with $T$ the temperature of the BH, is known to be determined by the local Rindler structure of the horizon \cite{Kiem:1995iy} and it  was proposed as a universal bound on the chaos that a quantum thermal system can develop \cite{Maldacena:2015waa}.   

    It this paper we compute the Lyapunov exponent for scattering near the photon-sphere of asymptotically AdS charged rotating BH's in $d=4$ and, later on, generalise our analysis to diverse dimensions $d$ and to horizon-free micro-state geometries a.k.a. {\it fuzzballs} \cite{Lunin:2002qf, Lunin:2001jy, Mathur:2005zp, Mathur:2008nj}.   
   We relate the Lyapunov exponent $\lambda$ to the vanishing rate of the radial velocity $\dot{r} \approx -2\,\lambda (r-r_c)$ near the photon-sphere, and derive analytic formulae for $\lambda$ as a function of the radius $r_c$ of the limiting photon orbit. 
   
We find that $\lambda$ is typically smaller than $\lambda_H$ characterising the vanishing rate of the radial velocity at the horizon\footnote{Our definition of $\lambda$ differs from that of \cite{Cardoso:2008bp} by a factor of 2. Our normalization is chosen so that $\lambda_H$ matches the Lyapunov exponent of the near horizon boundary theory.}, but the bound is violated inside a very narrow window near extremality when the innermost region of the BH photon-sphere coalesces with the BH horizon\footnote{The role of the ratio of $\lambda$, related to the minimum of the imaginary
part of the quasi-normal mode frequency, and $\lambda_H$, related to the surface gravity
of the inner horizon, based on \cite{Hintz:2015jkj} has been recently stressed in \cite{Hollands:2019whz} in connection with the strong cosmic censorship. We thank J.~Maldacena for pointing this out to us as well as the presence in this context of a sliver near extremality, similar to the one we find.}.      
 
We find\footnote{We have an analytical proof of this bound only for spherically symmetric BH's and numerical evidences for rotating black holes in four dimensions.} instead that 
\be
\lambda \leq \lambda_b = {\sqrt{d{-}3} \over 2 b_{\rm min}}
\ee
with $d$ the dimension and $ b_{\rm min}$ the minimal impact parameter under which a particle moving in the background of the BH gets trapped in the gravitational potential. The bound is saturated by non-rotating, uncharged BH's.

We remark that the existence of circular photon orbits and an instability timescale $\Delta t\approx 1/\lambda$ is a property common to many solutions in gravity, including BH-looking horizon-less geometries known as fuzzballs or BH microstate geometries with either flat or AdS asymptotics. A large class of horizon-free microstate geometries is known \cite{Bena:2015bea,Bena:2016agb,Bena:2016ypk} (and references therein) including their stringy origin \cite{Giusto:2009qq,Giusto:2011fy,Bianchi:2016bgx,Bianchi:2017bxl}. Geodesic motion in these geometries has been investigated in \cite{Bena:2017upb,Bena:2018mpb, Bena:2019azk, Bianchi:2018kzy, Bianchi:2017sds}. Here we focus on critical scattering for a class of three-charge BH microstate geometries with asymptotically flat or $AdS_3\times S^3$ geometries. We find that only geodesics scattered in asymptotically flat micro-state geometries can reach the fuzzball photon-sphere, exhibiting random angular dispersion at infinity. More interestingly, we find that the Lyapunov exponent $\lambda$ for a special class of asymptotically flat micro-state geometries is smaller than $\lambda_b$, which in turn is typically smaller than $\lambda_H$. Since $\lambda$ is related to the time decay of QNM's dominating at late times the response of the geometry to perturbations, our results suggest that ring-down signals can be used to discriminate between BH's and geometries with non trivial micro-structures at the horizon scale\footnote{Similar proposals were recently put forward in different contexts for the modifications of gravity at the horizon scale \cite{Cardoso:2019rvt, Barack:2018yly}.}.    
    
    The plan of the paper is as follows. In Section \ref{Geodesics} we discuss geodesics in asymptotically AdS Kerr-Newmann space-times in $d=4$. We derive the Hamiltonian for massless probes and discuss radial and angular motion by stressing the role of the impact parameter(s). We then consider geodesics ending inside the horizon. In section \ref{Critical} we study critical and nearly critical geodesics, we introduce the concepts of critical impact parameter $b_c$ and Lyapunov exponent $\lambda$ and propose a bound on the latter. In section \ref{Analytic} we discuss some BH examples in flat and AdS space-times. We also extend our analysis to arbitrary dimension $d$. In section \ref{Fuzzballs} we discuss three-charge fuzzball micro-state geometries in the near horizon limit or, more interestingly, in flat space-time. Section \ref{Conclusions} contains a discussion of our results and possible outlooks. In appendix \label{Angular} we discuss integration of the angular motion in $d=4$.

\section{ Geodesics on AdS Kerr-Newman space-times}
\label{Geodesics}

\subsection{The metric}

We consider geodetic motion in asymptotically $AdS_4$  Kerr-Newmann space-time\footnote{More precisely, the metric is not  asymptotically $AdS_4$ but rather tends to a rotating Einstein universe with $\Omega_\infty = -a/\ell^2$.} with metric \cite{Caldarelli:1999xj}  
\begin{equation}
\label{metric}
ds^2=-\frac{{\Delta_r}}{\rho ^2}\left(dt-\frac{a\sin^2\theta\,{d\phi}}{{{\alpha_\ell}} }\right)^2 + \frac{{\Delta_\theta}  \sin^2\theta}{\rho ^2}\left(a\,dt-\frac{\left(a^2+r^2\right) {d\phi}}{{{\alpha_\ell}} }\right)^2 + \frac{\rho ^2 dr^2}{{\Delta_r}} + \frac{\rho ^2 d\theta^2}{{\Delta_\theta} }
\end{equation}
where 
\begin{equation}
\begin{aligned}
\Delta_r &=\left(a^2+r^2\right) \left(\frac{r^2}{{{{\ell}}^2}}+1\right)-2 M r +Q^2\,,
\qquad
\rho^2 = r^2 + a^2\cos^2\theta  \,,
\\
\Delta_\theta&= 1 - \frac{a^2}{{{{\ell}}^2}}\cos^2\theta\,,
\qquad
{{\alpha_\ell}} =1 - \frac{a^2}{{{{\ell}}^2}}
\end{aligned}
\end{equation}
with $M$ the `mass', 
$a= J/M$ the angular momentum `parameter', ${{\ell}}$ the AdS radius and $Q$ the charge. A BH exists for any choice of $a<\ell$\footnote{This guarantees the positivity of 
the $g_{\theta\theta}$-component of the metric.}, $Q$ and $M$ such that $\Delta_r$ has at least a real positive root, and the horizon radius $r_H$ is then the largest root. 

A peculiar feature of the above metric is a non-zero angular velocity $\omega=g_{t\phi}/g_{tt}$ at infinity  $\Omega_\infty = {\omega}|_{ \infty} = - a /\ell^2$.  
The angular velocity that is relevant for the thermodynamics is the difference $\Omega = \Omega_H - \Omega_\infty$ with  $ \Omega_H ={\omega}|_{r_H}= a {{\alpha_\ell}}/(r_H^2 +a^2)$ the angular
velocity at the horizon \cite{Caldarelli:1999xj}.
 
 The Hawking temperature of the BH is given by 
\begin{equation}
2\pi \,T = {\Delta_r'(r_H)    \over 2(a^2+r_H^2)  } 
\end{equation}
If $\Delta_r'(r_H)=0$, the BH is said to be extremal and the temperature vanishes. The explicit expression of $r_H$ in terms of the BH parameters ($M$, $a$, $Q$ and the AdS `radius' $\ell$) can be found using Cardano-Tartaglia formulae for the zeroes of a quartic polynomial. We will refrain from exposing it since it is quite cumbersome and not crucial for our analysis.

\subsection{The Hamiltonian and momenta}

In the Hamilton-Jacobi formulation of the geodesics equation, the Hamiltonian reads
$${\mathcal H}=\ft12 g^{mn} P_m P_n$$ 
For a massless spin-less particle in the AdS Kerr-Newmann metric ${\mathcal H}$ can be written in the `separate' form
\begin{equation}
\label{hmltnn}
\begin{aligned}
\mathcal{H} &={1\over 2 \rho^2} \left(P_r^2 {\Delta_r}-\frac{\left( E (a^2+r^2)-a {{\alpha_\ell}}  {P_\phi} \right)^2}{{\Delta_r}}  \right)   + {1\over 2 \rho^2} \left( P_\theta^2 \Delta_\theta +\frac{(a E \sin^2\theta -{{\alpha_\ell}}  P_\phi)^2}{ \sin^2\theta\Delta_\theta }  \right)  
\end{aligned}
\end{equation}
with $E=-P_t$ and $P_\phi$ conserved quantities. Following Carter, Teukolsky and Chandrasekhar \cite{Chandrasekhar:1985kt}, the null condition $\mathcal{H}=0$ is solved by introducing a `separation' constant $K$, representing the total angular momentum, and setting 
 \begin{equation}
\label{Kdef}
\begin{aligned}
K^2 &=P_\theta^2 \Delta_\theta +\frac{(a E \sin^2\theta -{{\alpha_\ell}}  P_\phi)^2}{ \sin^2\theta\Delta_\theta }=- P_r^2 {\Delta_r}+\frac{\left(E (a^2+r^2)-a {{\alpha_\ell}}  {P_\phi} \right)^2}{\Delta_r}  
\end{aligned}
\end{equation}
Defining the impact parameters as
\be
{{\zeta}}={P_\phi {{\alpha_\ell}} \over E}-a    \qquad ,\qquad  b={K\over E} 
\label{var}
\ee
equations (\ref{Kdef}) can be written in the form
\begin{equation}
\begin{aligned}
 {{\cal R}}(r) &={  \Delta_r^2\, P_r^2\over E^2}  = \left(r^2  - a {{\zeta}} \right)^2-b^2\Delta_r 
\\
{{\varTheta}}(\cos\theta) &= {\Delta_\theta^2 P_\theta^2\,  \sin^2\theta\over E^2} =  b^2\,\Delta_\theta\,\sin^2\theta - (\zeta+a\, \cos^2\theta)^2 
  \label{p4}
\end{aligned}
\end{equation}
with ${{\cal R}}$ and ${{\varTheta}}$ quartic polynomials of $r$ and $\cos\theta$, respectively. 
While $b$ (as well as $a$ and $E$) can be taken to be non-negative without loss of generality, ${\zeta}$ can be negative or positive depending on $P_\phi$. One can distinguish three classes
of geodesics according to the sign of $P_\phi$:
\be
\begin{aligned}
{{\zeta}}>-a && \qquad  \text{co-rotating} \\
{{\zeta}}=-a && \qquad  \text{non-rotating} \\
{{\zeta}}<-a && \qquad  \text{counter-rotating}
\end{aligned}
\ee  
We notice that at infinity
\be
 {{\cal R}}(r)  \approx  r^4\left(1 -{b^2 \over \ell^2} \right)+O(r^2)
\ee
so positivity of $P_r^2$ requires\footnote{One can also consider geodesics that never reach infinity and allow for $b>\ell$. We will comment on this case later on.} 
\be
b< \ell    \label{bell} 
\ee 
In fact we will later see that $b$ is also bounded from below for the motion to take place at all. Using the monotonically decreasing variable $r$  as time variable in the branch where the geodesics evolves towards the BH, so that $\dot{r},P_r< 0$, the equations of motion $\dot{x}^m={ \partial {\cal H}/ \partial P_m}$  can be written in the form
\begin{equation}
\begin{aligned}
\frac{dt}{dr} &= {E\over \Delta_r P_r}\left[ {a  \left(  {{\zeta}}+a  \cos^2\theta  \right)\over \Delta_\theta}+{(r^2+a^2)\left( r^2-a\,{{\zeta}}    \right)\over \Delta_r}\right] \\
\frac{d\phi}{dr} &= {E\, {{\alpha_\ell}}\over \Delta_r P_r}\left[ { \left({{\zeta}} +a  \cos^2\theta  \right)\over \sin^2\theta\, \Delta_\theta}+{ a\, \left(r^2- a\,{{\zeta}}    \right)\over \Delta_r}\right] \\
\frac{d\theta}{d r} & = \frac{\Delta_\theta P_\theta}{\Delta_r P_r} \label{velocities}
\end{aligned}
\end{equation}

\subsection{Motion in the radial direction}

\begin{figure}[t]
\centering
 \includegraphics[scale=0.45]{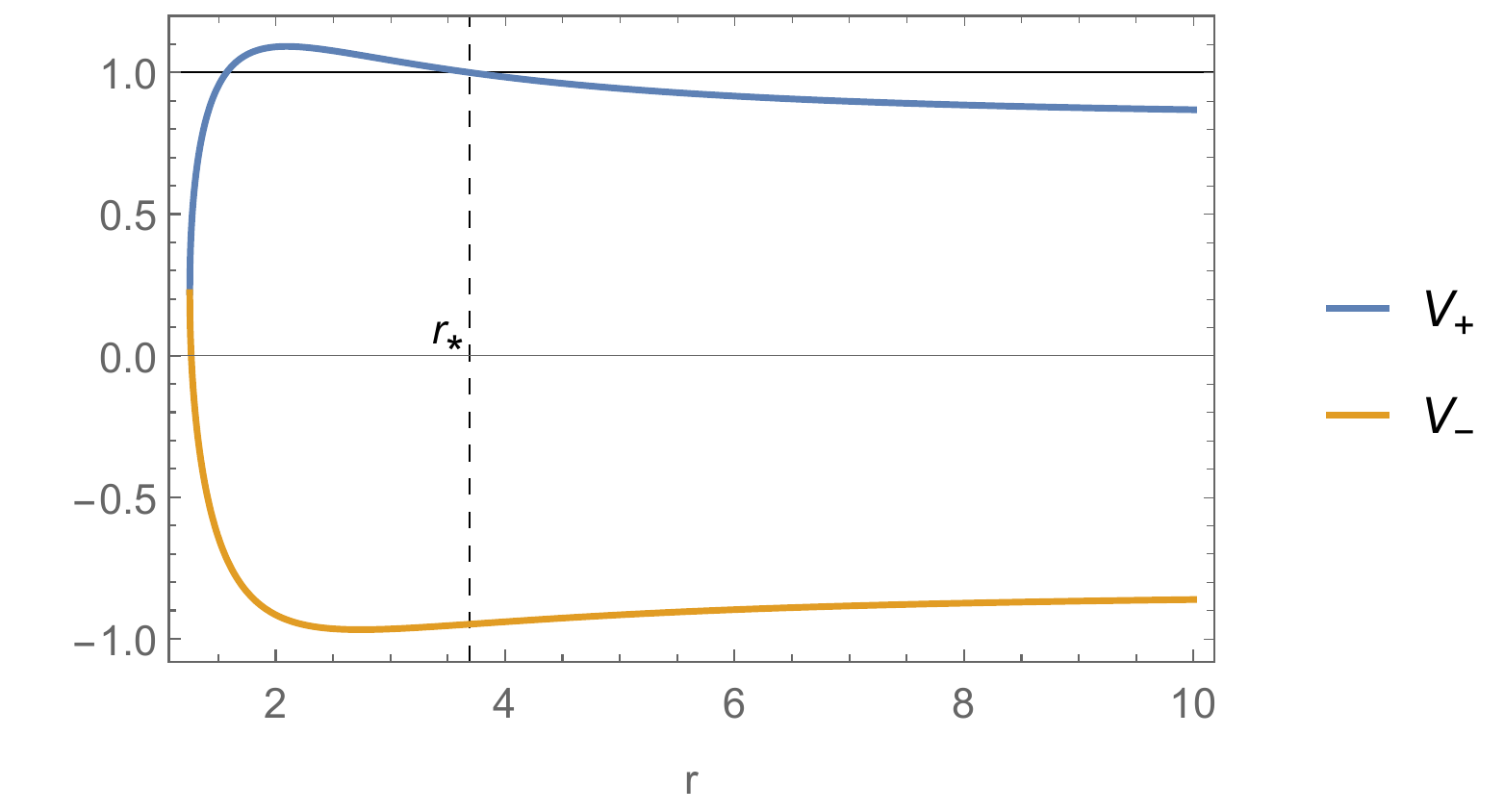}\quad
  \includegraphics[scale=0.45]{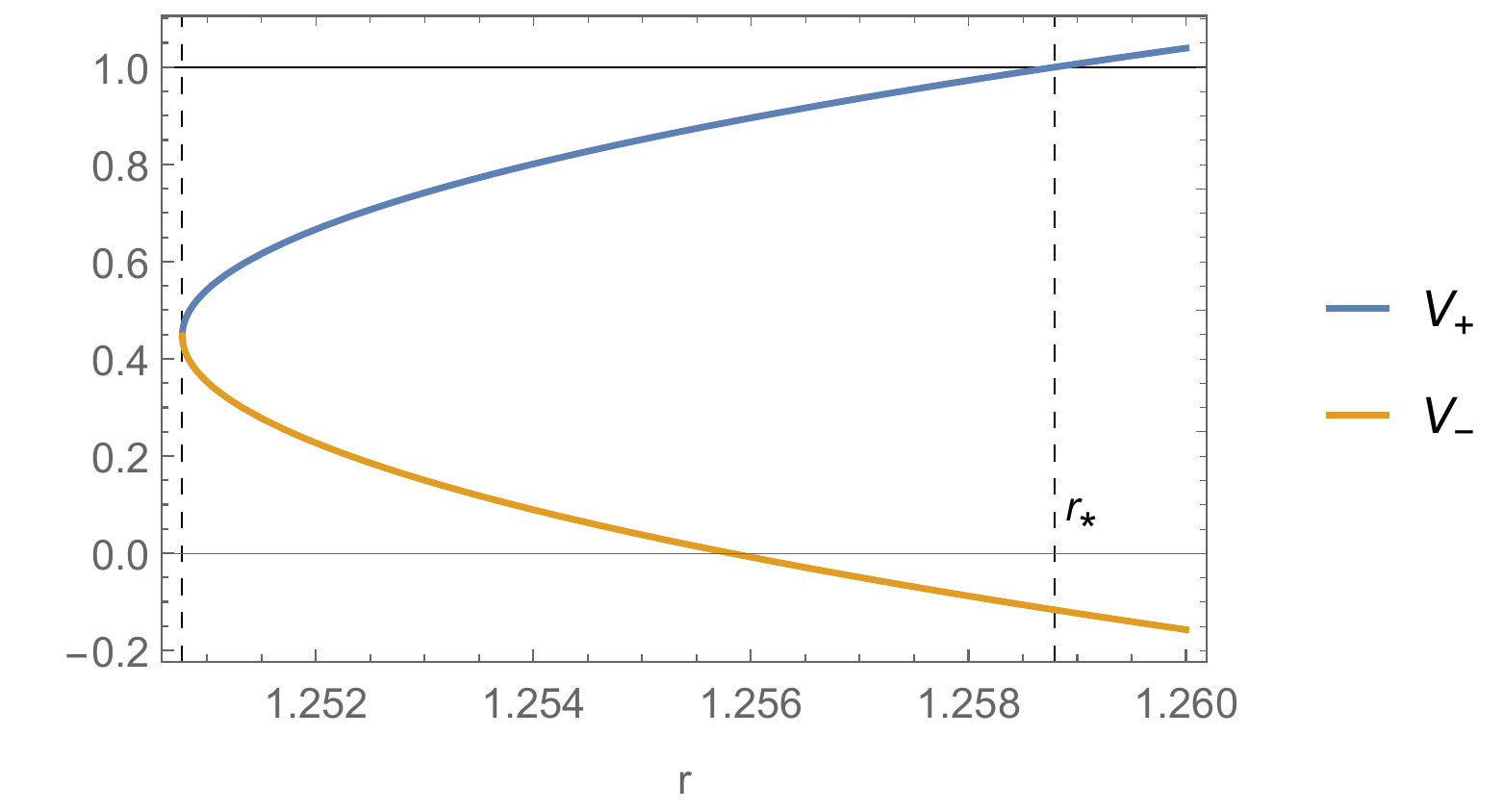}
\caption{\footnotesize{Effective potentials  $V_\pm(r)$ for a BH with: $M=1$, $\ell=3$, $a=0.7$, $Q=0.3$ and impact parameters $\zeta=0.5$ a) $b=2.5$: The particle reaches the turning point $r_*$ and returns to infinity b) $b=8>\ell$. Motion is allowed only inside a finite region of the space and geodesics always fall into the horizon. }}
\label{fvpvm}
\end{figure}

We can view the motion in the radial direction as that of a particle in a one-dimensional potential. We write
\be
{ {\cal R}(r) \over r^4} =\left[1-V_+(r) \right]\left[1-V_-(r) \right]  
\ee
with $V_\pm$ the effective potentials
\be
  V_\pm(r)= {a\, \zeta\over r^2} \pm {b \sqrt{\Delta_r(r)}\over r^2 }
\ee
Starting from infinity, a massless particle with large enough impact parameter $b$ moves towards the BH till it   
 reaches a turning point $r_*$,  where $V_+(r_*)=1$ and the radial velocity vanishes, see figure \ref{fvpvm}. 
  After this point, the particle bounces back to infinity along a symmetrically reversed trajectory obtained by flipping the sign of the radial velocity.

Critical geodesics are obtained by tuning the impact parameters so that the value of $V_+(r)$ at the maximum is one, i.e.  
\be
V_+'(r_c)=0 \qquad {\rm and} \qquad   V_+(r_c)=1   
\ee
for some $r_c>r_H$. We will see that for $b$ smaller than a critical value $b_c(\zeta_c)$, the potential satisfies $V_+(r)<1$ for all $r>r_H$ and the geodesics finds no turning point. So, we can divide
the impact parameter plane $(b,\zeta)$ into three regions
\bea
b<b_c(\zeta_c)&& \qquad ,\qquad {\rm absorption~phase}\nn\\
b_c(\zeta_c)<b <\ell && \qquad ,\qquad {\rm scattering~phase} \nn\\
b>\ell && \qquad ,\qquad {\rm only~motion~in~the~AdS~interior}
\eea   
    We will mainly focus on the second class of geodesics.
  
\subsection{Motion in the $\theta$ direction}
\label{ThetaMotion}

Motion along the $\theta$ direction takes place inside intervals bounded by the zeros of the quartic polynomial  ${{\varTheta}}(\cos\theta)$. 
Denoting $\chi=\cos\theta$,  the polynomial  ${{\varTheta}}$ in (\ref{p4})  can be conveniently written in the form
 \bea
 {{\varTheta}}(\chi)&=&  b^2 \, \left( 1-{a^2\chi^2\over \ell^2} \right) (1-\chi^2)  - \left( \zeta+a\chi^2 \right)^2 \nn\\
 &=& A \, \chi^4+B \,\chi^2+C=A\,(\chi^2 -\chi_{\rm p}^2)(\chi^2 -\chi_{\rm m}^2)  \label{thetapol}
 \eea
with
\begin{equation}
\begin{aligned}
A &= -a^2 \left( 1-{b^2\over \ell^2} \right)  ~,~ \quad
B &=  -b^2  \left( 1+{a^2\over \ell^2} \right)-2\,a\,{{\zeta}} ~,~\quad C = b^2-{{\zeta}}^2 
\end{aligned}
\end{equation}

The zeros of the polynomial ${{\varTheta}}(\chi)$ are located at 
\be
\label{chipm}
\chi^2_{\rm p,m }  = {-b^2  \left(  1+{a^2\over \ell^2} \right)-2 \,a\, {{\zeta}} \mp 
   \sqrt{ \Delta}\over 2\, a^2\, \left(1-{b^2\over \ell^2 }\right)} 
\ee
with
\be
     \Delta = b^2\left[ b^2 \left( 1+{a^2\over \ell^2} \right)^2 + 4a^2 \left( 1+ { \zeta^2\over \ell^2} \right)+
 4a\zeta \left( 1+{a^2\over \ell^2} \right) \right] \: .
\ee

Geodetic motion is allowed  for choices of the parameters $(b,\zeta)$ such that $P_\theta^2>0$ for some $\chi \in \left[-1,1\right]$. 
To determine this region it is convenient to write $P_\theta$ in the following form 
\be
{P_\theta^2 \Delta_\theta \over E^2} =b^2- V_{\rm eff}(\chi) 
\ee 
with the angular effective potential
\be
V_{\rm eff}(\chi)= { \left( \zeta+a\chi^2 \right)^2 \over \left( 1-{a^2\chi^2\over \ell^2} \right) (1-\chi^2) }
\ee
The effective potential is even in $\chi$ and bounded from below inside the interval $\left[-1,1\right]$. For $\zeta=-a$ it has a maximum at $\chi_*=0$,
and minima at $\chi_*=\pm 1$. For $\zeta \neq -a$ it reaches infinity at $\chi=\pm 1$.  
 The extrema of the potential (solutions of $V'_{\rm eff}(\chi_*)=0$) are located at $\chi=\pm \chi_*$ (if real and with $\chi_*^2\leq 1$) with
\be
\chi_*^2=\left\{ 0, -{\zeta\over a} , { \ell^2 (\zeta-\zeta_*)\over a(\ell^2 -\zeta \, \zeta_*)  }   \right\}  \qquad {\rm with} \qquad   \zeta^* =- \frac{2a\ell^2}{\ell^2+a^2} \label{3min}
\ee
and
\be
V_{\rm eff} (\chi_*)= \left\{\zeta^2, 0,  -{4 a \ell^2 (\zeta+a)(\ell^2+a \zeta) \over (\ell^2-a^2)^2} \right\}  
\ee
\begin{figure}[t]
\centering
\includegraphics[scale=0.5]{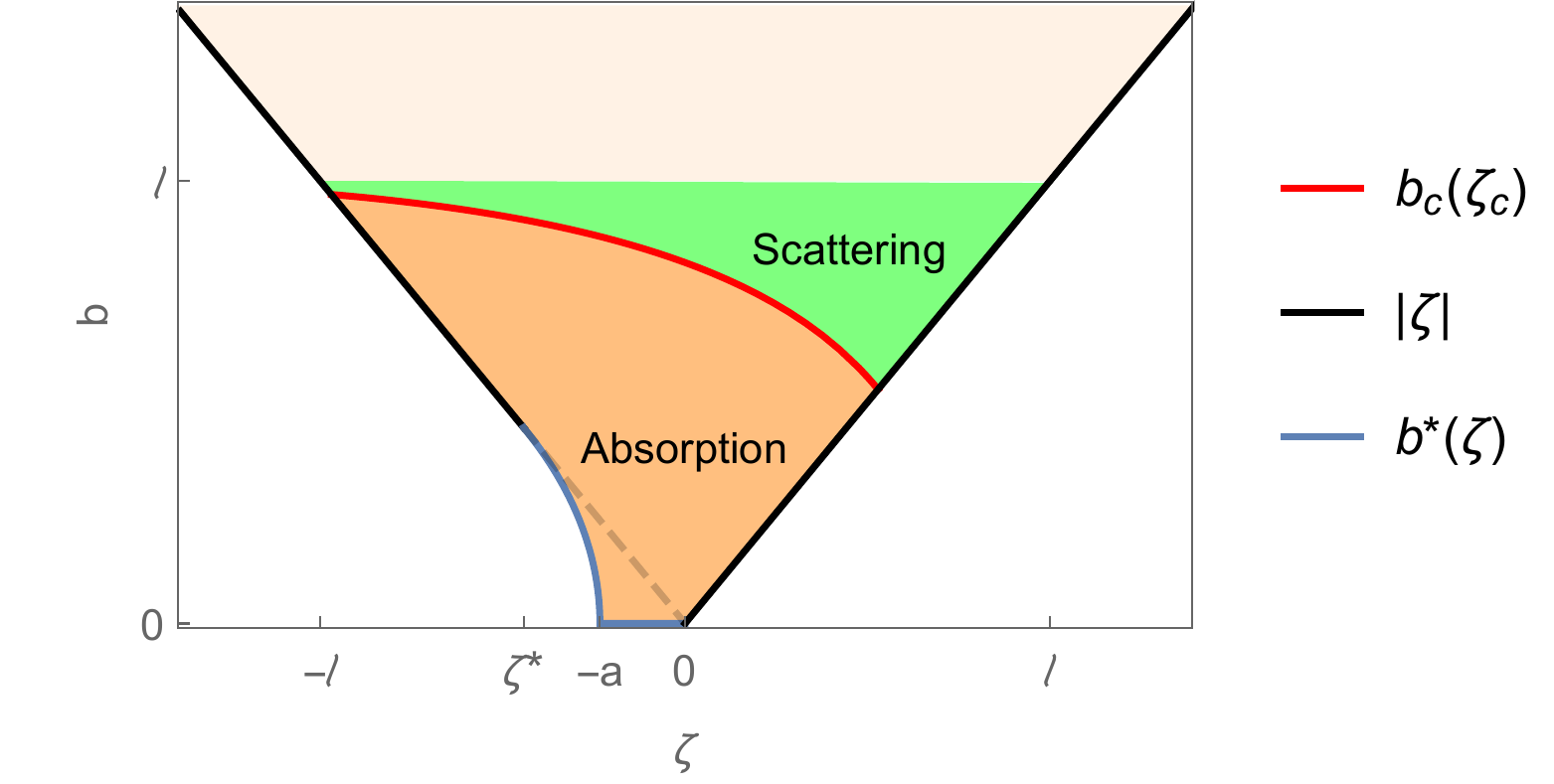}
\caption{\footnotesize{Allowed regions (green, red and light red) in the $\zeta$, $b$ plane of a BH with parameters: $M = 1$, $\ell=3$, $a=0.7$. Geodesics in the light red region does not reach infinity. }}
\label{regions}
\end{figure}
 We are interested on geodesics starting from infinity, so we will  take $b^2<\ell^2$. In addition $b^2$ should be bigger than the absolute minimum of $V_{\rm eff} (\chi_*)$ inside the interval  $\chi \in \left[-1,1\right]$. 
 Depending on the value of $\zeta$, the profile of $V_{\rm eff} (\chi)$ changes and the global minimum can be located at any of the three extrema in (\ref{3min}).  
 We have  two different allowed regions:

\begin{enumerate}[A)]
\item
The geodesics bounces inside the interval $\chi\in  \left[-\chi_{\rm p},\chi_{\rm p}\right]$ for
\be
 -\ell < \zeta< \ell \quad {\rm and} \quad |\zeta|<b<\ell   
 \ee
 For $\zeta_*<\zeta<0$, the effective potential has two minima and a maximum in $\chi=0$, while outside this interval the potential has a unique minimum in $\chi=0$.
\item
The geodesics bounces inside the interval $\chi\in    \left[-\chi_{\rm p},-\chi_{\rm m}\right]$ or $ \chi\in  \left[\chi_{\rm m},-\chi_{\rm m}\right]$ 
\be
 \zeta_* < \zeta < 0 \quad {\rm and}  \quad b_*(\zeta)<b<|\zeta|  
 \ee
 with 
  \be
b_*^2(\zeta)= \begin{cases} 
-{4 a \ell^2 (\zeta+a)(\ell^2+a \zeta) \over (\ell^2-a^2)^2}\quad&\text{for}\quad \zeta_*<\zeta<-a
\\
0\quad&\text{for}\quad -a<\zeta<0
\end{cases}
\ee  
  The effective potential has two minima and a maximum in $\chi=0$. 
  
\end{enumerate}

We display in colors the allowed region in the impact parameter plane $(b,\zeta)$ in figure \ref{regions}. 
Geodesics belonging to the classes A and B correspond to the regions above and below $b=|\zeta|$ (dashed line in the graph) respectively.
We distinguish with two colors, red and green,  the absorption and scattering phases  respectively (see below for details).   The light red region collects all geodesics 
allowed by the kinematics that cannot reach the AdS boundary at infinity. Geodesics in this class always fall into the horizon.   
  
\begin{figure}[t]
\centering
\subfloat[]{\includegraphics[scale=0.5]{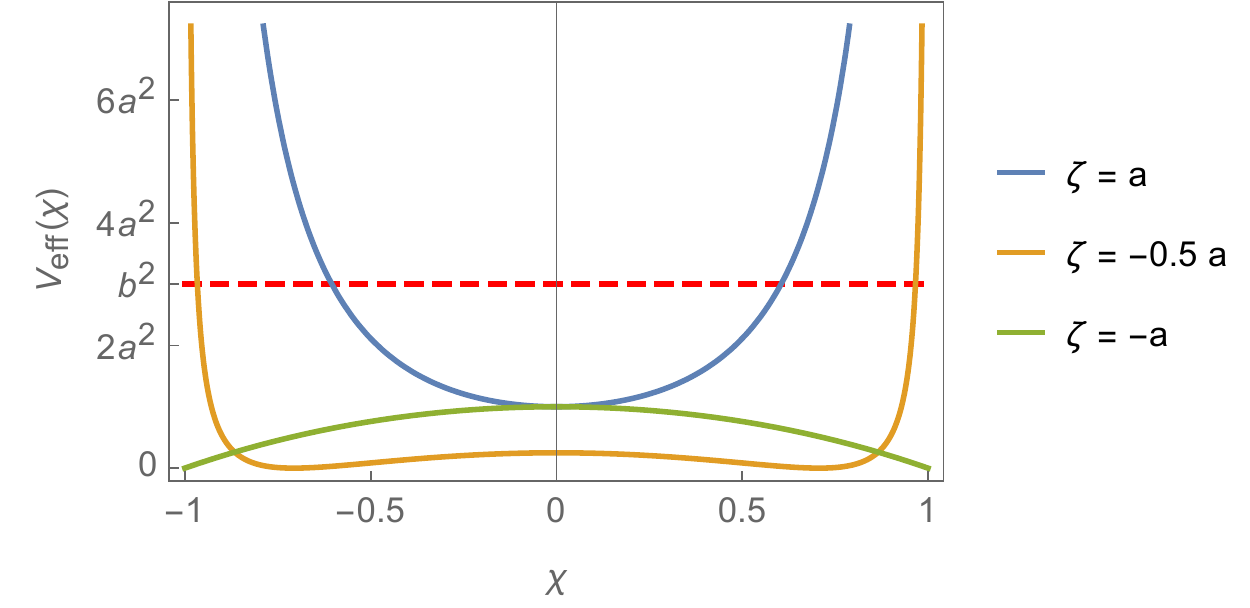}}
\quad
\subfloat[]{\includegraphics[scale=0.5]{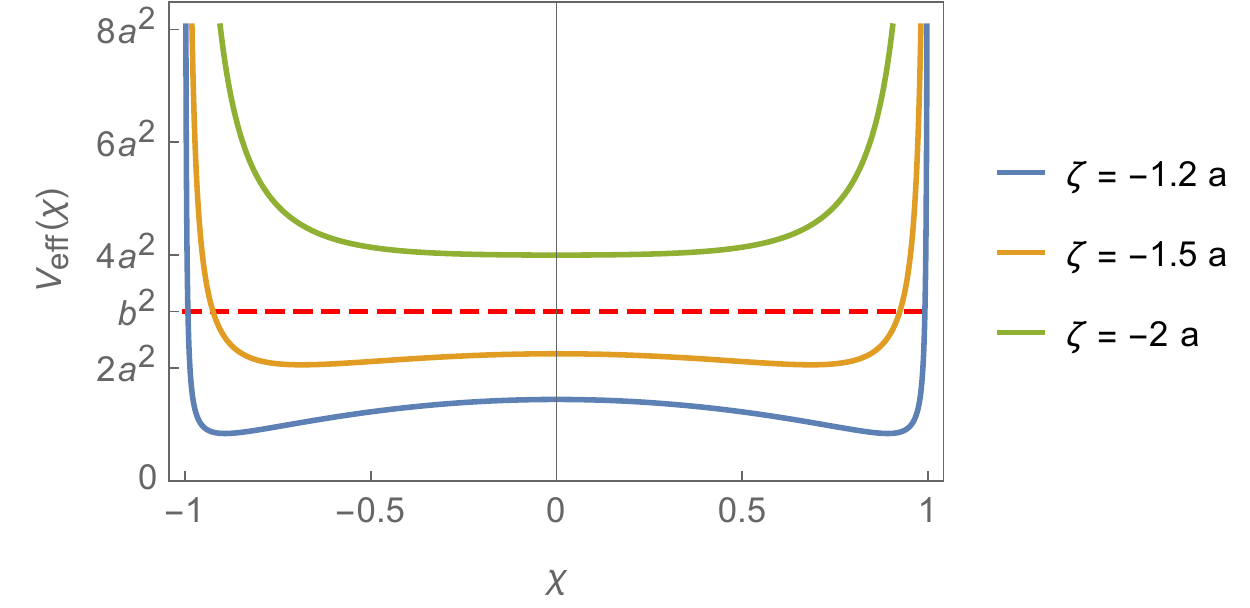}}
\caption{\footnotesize{Effective potential $V_{\rm eff}(\chi)$ for a BH with angular parameter $a=0.7$ , $\ell=3$: (a)  Co-rotating and non-rotating geodesics (b)  Counter-rotating geodesics.}}
\label{vthetaplot}
\end{figure}

In figure \ref{vthetaplot}, we display the form of $V_{\rm eff}(\chi)$ for various choices of $\zeta$  setting $a=1$, $\ell=2$, $Q=0.3$. Geodetic motion is allowed when $V_{\rm eff}(\chi)$ is below the $b^2$-dashed line and it bounces inside the intervals limited by the intersection points of the two graphs. 
Whenever $b^2$ coincides with an extremum of $ V_{\rm eff}(\chi)$ one has either a shear-free (un)stable geodesics at fixed $\theta=\theta_{0}$ or a critical geodesics at the equator 
 (see next section). Geodesics at fixed $\theta$-angle are stable around the minima of $ V_{\rm eff}(\chi)$, and unstable around the maximum. 
 
We will see in the next section that critical geodesics always correspond to the case $b>|\zeta|$, so critical motion will always bounce inside the interval $\chi \in \left[ -\chi_+,\chi_+\right]$
with $\chi_+=1$ for $\zeta=-a$. 

 \subsection{Geodesics inside the absorption region}
\label{geoEndBHhor}

 For $b<b_c(\zeta_c)$, the particle falls inside the BH horizon. It is important to observe that near the horizon both radial and angular velocities along the $\phi$-direction are independent
 of the impact parameters $(b,\zeta)$. Indeed expanding  (\ref{velocities}) near $r\approx r_H$ one finds ${\cal R} \approx (r_H^2-a \, \zeta)^2$ leading to
\begin{equation}
\begin{aligned}
\frac{dr}{dt} &\approx { \Delta_r'(r_H) (r-r_H) \over   (a^2+r_H^2) }  \quad, \quad
\frac{dr}{d\phi} \approx { \Delta_r'(r_H) (r-r_H) \over   {{\alpha_\ell}}\, a } \quad, \quad
\frac{dr}{d \chi}  \approx -\frac{  (r_H^2-a \, \zeta)  }{ \sqrt{{\varTheta}(\chi)} } \label{velocitiesrh}
\end{aligned}
\end{equation}
 We notice that near the horizon both the distance from the horizon and the radial velocity vanish but their ratio
   \begin{equation}
\begin{aligned}
\lambda_H=-{1\over 2(r-r_H)} \frac{dr}{dt} \Big|_{r_H} \approx  {\Delta_r'(r_H)    \over 2(a^2+r_H^2)  } = 2\pi T
\end{aligned}
\end{equation}
is finite and proportional to the BH temperature. We will show later that $\lambda_H$ is typically bigger than the ratio $\lambda$ describing the vanishing rate of the radial velocity at the photon-sphere. This inequality is however violated for extremal and near-extremal black holes.  
 
 \subsection*{Shear free geodesics}
 
   A particular simple class of geodesics in the absorption phase are the so called shear-free geodesics (radial falling) corresponding to trajectories with zero total angular momentum. 
For $b=0$, non-negativity of $P^2_\theta$ requires that the polynomial ${\varTheta}(\cos\theta)$ in (\ref{p4}) exactly vanishes. This happens when the impact parameter $\zeta$ and the initial angle $\theta_0$ 
are related by 
   \be
  {{\zeta}}=-a\cos^2\theta_0 
   \ee
For this choice $\dot\theta=0$, i.e. $\theta=\theta_0$ along the whole trajectory.
Plugging these values into the polynomial ${\cal R}(r)$ one finds
    \be
    {\cal R}(r)=(r^2+a^2 \,\cos^2\theta_0)^2
    \ee
that  is strictly positive, so no turning points are found before the geodesics reaches the BH horizon.   
The rates of change of $t$, $\phi$ and $\theta$ with $r$ take the simple form
       \begin{equation}
\begin{aligned}
\frac{dt}{dr} &=   {a^2+r^2 \over \Delta_r(r)  }   \quad, \quad \frac{d\phi}{dr} =   {{{\alpha_\ell}}\, a\,  \over \Delta_r(r) } \quad, 
\quad \frac{d\theta}{d r}  = 0 \label{velocities2}
\end{aligned}
\end{equation}
that reduce to (\ref{velocitiesrh}) near the horizon.

\section{Critical and nearly critical geodesics}
\label{Critical}

In this section we study critical and nearly critical geodesics in asymptotically AdS Kerr-Newman BH's. 

\subsection{Critical geodesics}

A critical geodesics is obtained when the two largest roots 
of the polynomial
\begin{equation}
{{\cal R}}(r) =  \left( r^2  - a\,{{\zeta}}  \right)^2- b^2\, \Delta_r
\end{equation}
coincide. 
  More precisely,  we say that a geodesics is critical, if the largest zero $r_c$ of  ${{\cal R}}(r)$ is a double zero, or in other words
 \be
 {{\cal R}}(r_c)={{\cal R}}'(r_c)=0\label{critical}
 \ee
  These equations can be easily solved for ${{\zeta}}$ and $b$ as a function of the critical radius
\bea
 b_c(r_c) =  {4 \, r_c \, \sqrt{\Delta_r(r_c)}\over \Delta_r'(r_c) }\qquad , \qquad 
  {{\zeta}}_c(r_c)  = {r_c^2\over a} -  {     4r_c \Delta_r(r_c)   \over a \Delta'_r(r_c) }     
\label{xic} 
\eea
\begin{figure}[t]
\subfloat[]{\includegraphics[scale=0.45]{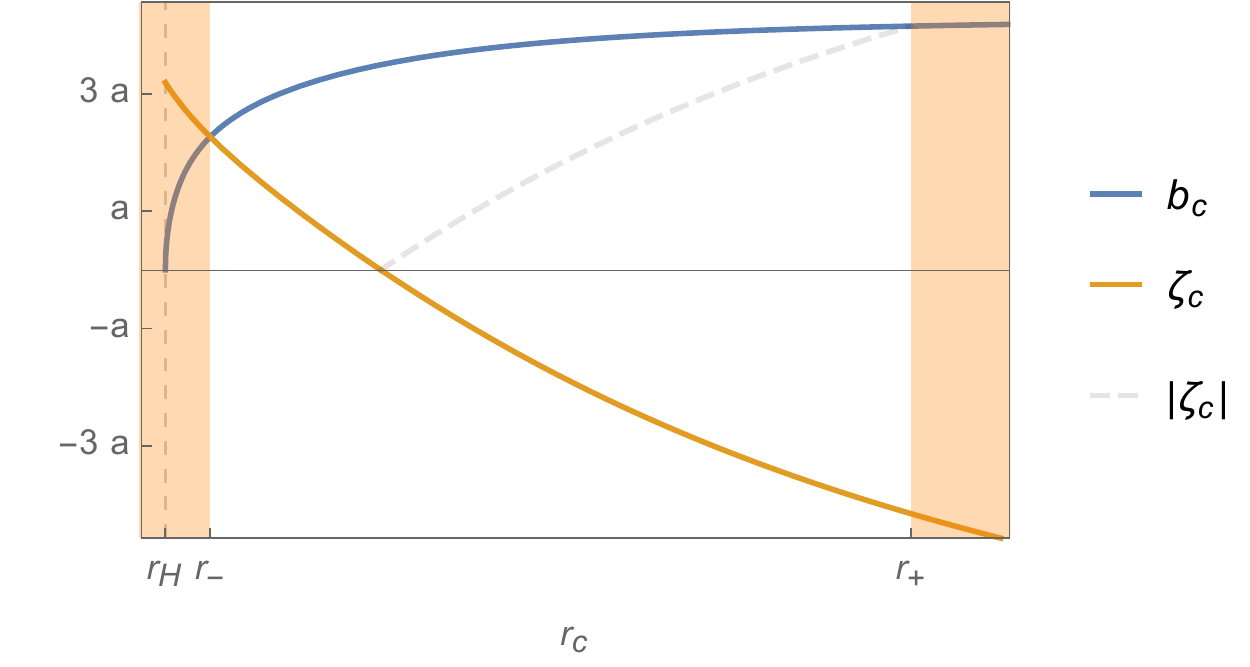}}
\qquad
\subfloat[]{\includegraphics[scale=0.45]{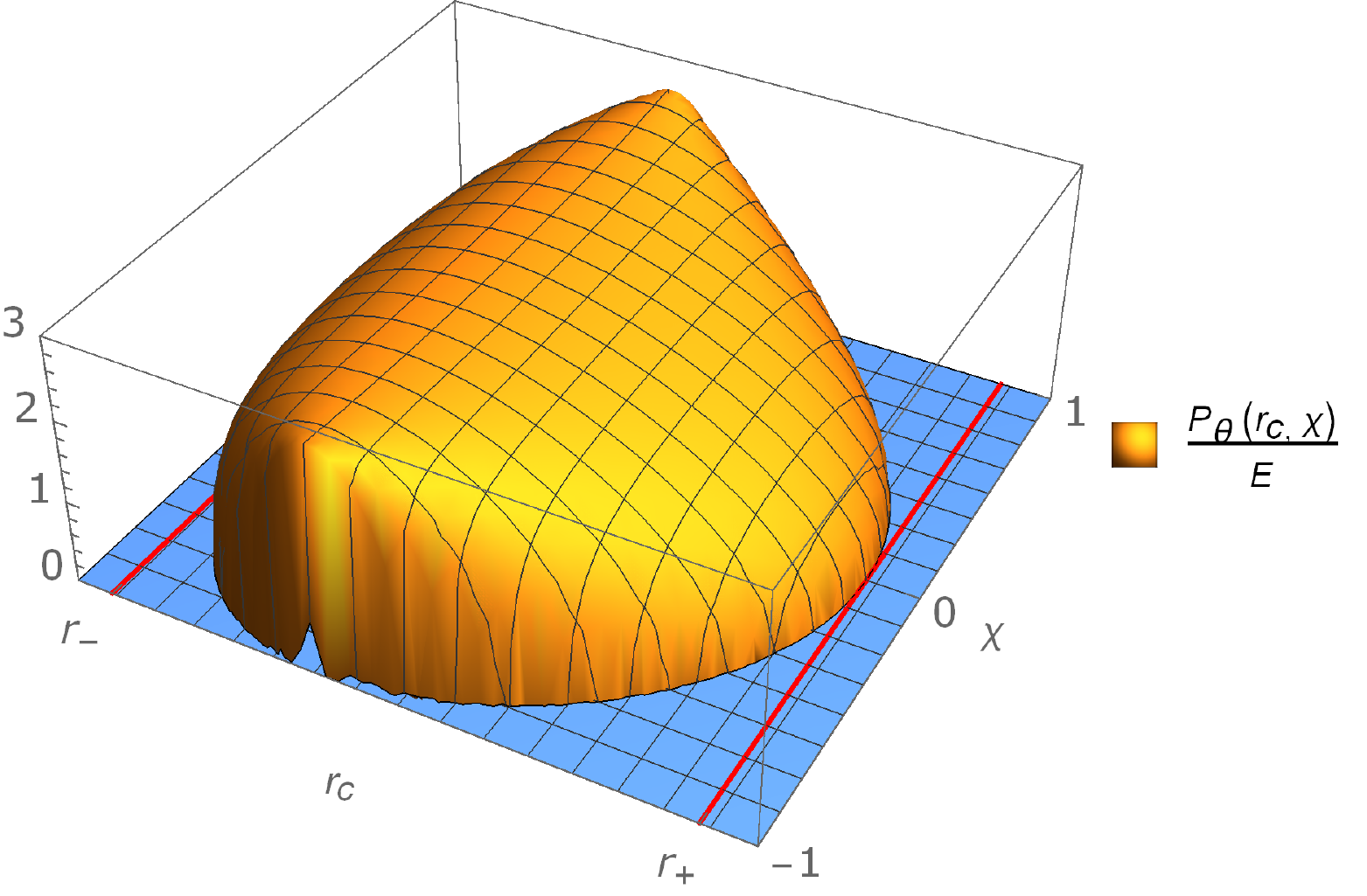}}
\caption{\footnotesize{(a) Plots of $b_c(r_c)$ and ${{\zeta}}_c(r_c)$. (b) Plot of $P_\theta(r_c,\chi)$ for AdS Kerr-Newman BH's  with mass $M=1$, AdS radius $\ell = 3$, angular momentum $a=0.7$ and charge $Q=0.3$. }}
\label{thethaplots}
\end{figure}

Plugging these formulae into ${\varTheta}(\chi)$, one finds that the polynomial is positive  for $r_c$ belonging to a finite interval 
\be
r_c \in  \left[ r_{-},r_{+}\right] \label{interval}
\ee
wherein
\be
b_c(r_c )^2 \geq   {{\zeta}}_c(r_c )^2 \label{bineq}
\ee 
  with $r_\pm$ saturating the inequality. 
   We remark that for non-extremal BH's , $r_-$ is  always outside the horizon since equations (\ref{critical}) do not admit any solution with $r_c=r_H$ 
 unless $\Delta'(r_H)=0$. On the other hand, one can see that depending on the masses, angular momentum and charges, $r_-$ can be either inside or outside the ergo-region. 
    
  Geodesic saturating the inequality in (\ref{bineq}) lie on the equatorial plane $\chi=\dot\chi=0$ and their impact parameters take the simple form 
    \be
   b_\pm =\mp {{\zeta}}_\pm=  {4 \, r_\pm \, \sqrt{\Delta_r(r_\pm)} \over \Delta_r'(r_\pm) }  \label{zetapm}
  \ee
  where ${{\zeta}}_\pm={{\zeta}}(r_\pm)$ and  $b_\pm=b_c(r_\pm)$.

    On the other hand, critical geodesics with $r_c$ inside the interval $\left[ r_-,r_+ \right]$ display a non-trivial motion in the $\theta$ direction. 
    In figure  \ref{thethaplots} we display the dependence of   $({{\zeta}}_c,b_c)$ and $P_\theta$ on the critical radius $r_c$.
  We notice that in the region where $P_\theta^2$ is positive and therefore motion is allowed $b_c>|\zeta_c|$, so critical geodesics fall always into the first of the two classes identified in Section 2.3. The motion in the $\chi$-plane is confined inside the interval $\left[-\chi_+, \chi_+\right]$.

\begin{figure}[t]
\centering
\subfloat[]{\includegraphics[scale=0.3]{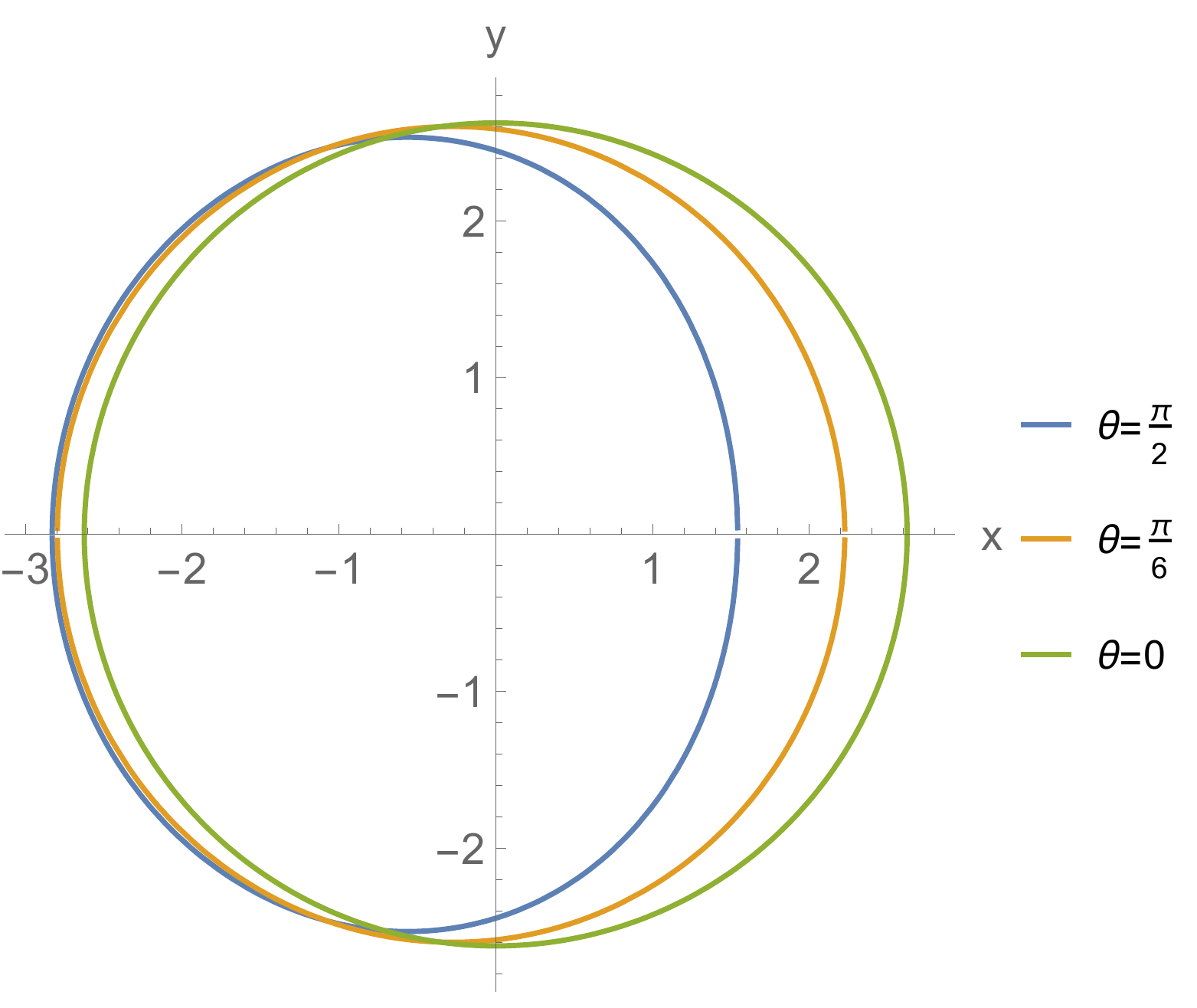}}
\caption{\footnotesize{(a) Shadow of a BH in AdS space with parameters $\ell=3$, $M=1$, $a=0.7$, $Q=0.3$ as seen by an observer with inclination $\theta = 0$, $\theta=\pi/6$ and $\theta=\pi/2$ (equatorial plane).}}
\label{fshadow}
\end{figure}

  The collection of all points leading to a critical geodesics delimits the edge of the BH shadow 
  as seen by an observer at infinity. For an observer on a plane at fixed $\theta$, the rim of the shadow is defined by the parametric curve \cite{Chandrasekhar:1985kt} 
   \be
   (x- x_0)^2 + y^2 = b_c^2 \label{rimshad}
  \ee
 where 
  \be
  y={P_\theta (r_c) \sqrt{\Delta}_\theta \over E}~,~ \quad 
  x={{{\alpha_\ell}} P_\phi(r_c) \over E \sin\theta \sqrt{\Delta}_\theta}~,~ \quad x_0 = {a \sin\theta\over \sqrt{\Delta}_\theta} 
  \ee
   with $r_c$ running inside the sub-interval of (\ref{interval}) allowed by the chosen value of $\chi=\cos\theta$, see Fig. \ref{thethaplots}. 
  In figure \ref{fshadow} we depict the shadow of a Kerr-Newmann AdS BH for particular values of mass $M$, charge $Q$ and angular momentum parameter $a$ at different values of $\theta=0,\pi/6,\pi/2$. Note the asymmetric shape, due to the rotation, as one approaches the equatorial plane $\theta=\pi/2$.

\subsection{The Lyapunov exponent}

In this section we consider nearly critical geodesics along the equatorial plane  ($\theta=\pi/2$, $\chi=0$) and compute the Lyapunov exponent determining the exponential growth with time of geodetic deviation.  

Approaching the critical radius $r\approx r_c$ with $\chi=\dot{\chi}=0$, the radial velocity (\ref{velocities})  vanishes as
\be
{dr\over dt} \approx - 2\, \lambda \, (r-r_c) \label{radial0}
\ee
with 
\begin{equation}
\begin{aligned}\label{lambda}
    \lambda &=      \ft12  \sqrt{ {\cal R}''(r_c) \over 2}   {  \Delta_r(r_c)    \over  a  {{\zeta}}_c  \Delta_r(r_c)  
     +\left( r_c^2 -a\,{{\zeta}}_c   \right)(a^2+r_c^2)      }\end{aligned}
\end{equation}
 and
    \bea
   {\cal R}''(r_c) = 12\, r_c^2 -4\,  a\,  {{\zeta}}_c -b_c^2   \Delta_r''(r_c)   
   \eea
    We will now show that $\lambda$, governing the vanishing rate of $\dot{r}$, coincides with the Lyapunov exponent of nearly critical geodesics scattered around $r\approx r_c$. 
     A nearly critical geodesics can be obtained by varying the impact parameters $(b,{{\zeta}})$ slightly away from the critical values $(b_c,{{\zeta}}_c)$. There are two distinct 
cases depending on whether the variation lifts the double zero or splits it into two. In the former case, the turning point disappears and the (massless) particle falls into the BH. We are interested in the latter 
case whereby the (massless) particle reaches the largest of the two nearby zeroes and bounces  back to infinity. 
Near the critical point the polynomial ${\cal R}(r)$ can be approximated by
 \be
 {\cal R}(r) \approx c \left[  (r-r_c)^2 -    \epsilon^2 \right]=c (r-r_c-  \epsilon)(r-r_c+  \epsilon)
 \ee
 where $\epsilon$ parametrises the distance between the critical radius and the turning point of the nearly critical geodesics located at $r_*=r_c+  \epsilon$. The time delay to reach $r_*$ and get back becomes
 \begin{equation}
\begin{aligned}
\Delta t & =2\int^{r_*} {dr \over \dot{r} } \approx -   {1\over \lambda} \int^{r_*}
{dr \over    \sqrt{(r-r_*)(r-r_c+  \epsilon) }   } 
 \approx  - {1\over \lambda} \log \epsilon
\end{aligned}
\end{equation}
 where (\ref{radial0}) has been used. 
 Similarly for the deflection angle one finds
  \begin{equation}
\begin{aligned}
\Delta \phi &= 2\,  \,\int^{r_*} {\dot{\phi} \, dr \over \dot{r} }  \approx -  {{\omega_c} \over \lambda} \, \int^{r_*}
{  dr \over   \sqrt{(r-r_*)(r-r_c+ \epsilon) }   } 
 \approx   -{{\omega_c} \over \lambda}\, \log \epsilon
\end{aligned}
\end{equation}
 with 
 \be
 {\omega_c}=\dot{\phi}(r_c)= {{\alpha_\ell}}   \left[{ 
  {{\zeta}}_c   \Delta_r(r_c)    + a\, \left(r_c^2- a\,{{\zeta}}_c    \right)  \over  a    {{\zeta}}_c  \Delta_r(r_c)  +(r_c^2+a^2)\left( r_c^2-a\,{{\zeta}}_c    \right)        }\right]
 \ee
  the angular velocity at the critical point. 
  Taking the variation with respect to the deviation in the incoming momenta one finds 
 \be
{ \delta (\Delta\phi)\over \delta P_\phi }\sim { \delta (\Delta\phi)\over \delta \epsilon}\approx {{\omega_c} \over \lambda\, \epsilon}\approx {{\omega_c} \over \lambda}\,  e^{ \lambda \, \Delta t}
 \ee
   We  conclude that the Lyapunov exponent $\lambda$ is determined by the vanishing rate (\ref{radial0}) of the radial velocity of an infalling neutral massless particle at the photon-sphere.

  \begin{figure}[t]
\centering
\subfloat[]{\includegraphics[scale=0.45]{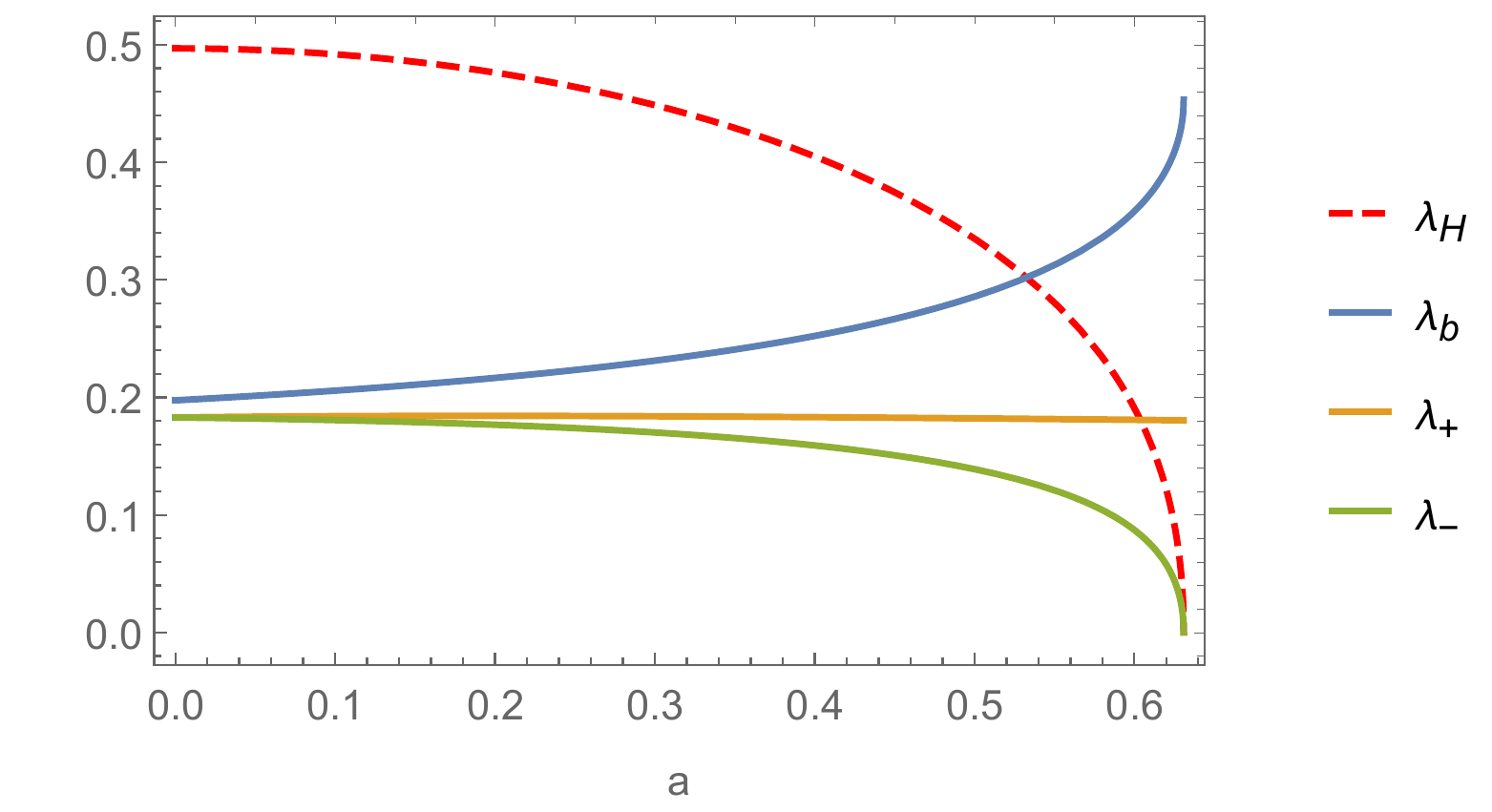}}
\qquad
\subfloat[]{\includegraphics[scale=0.45]{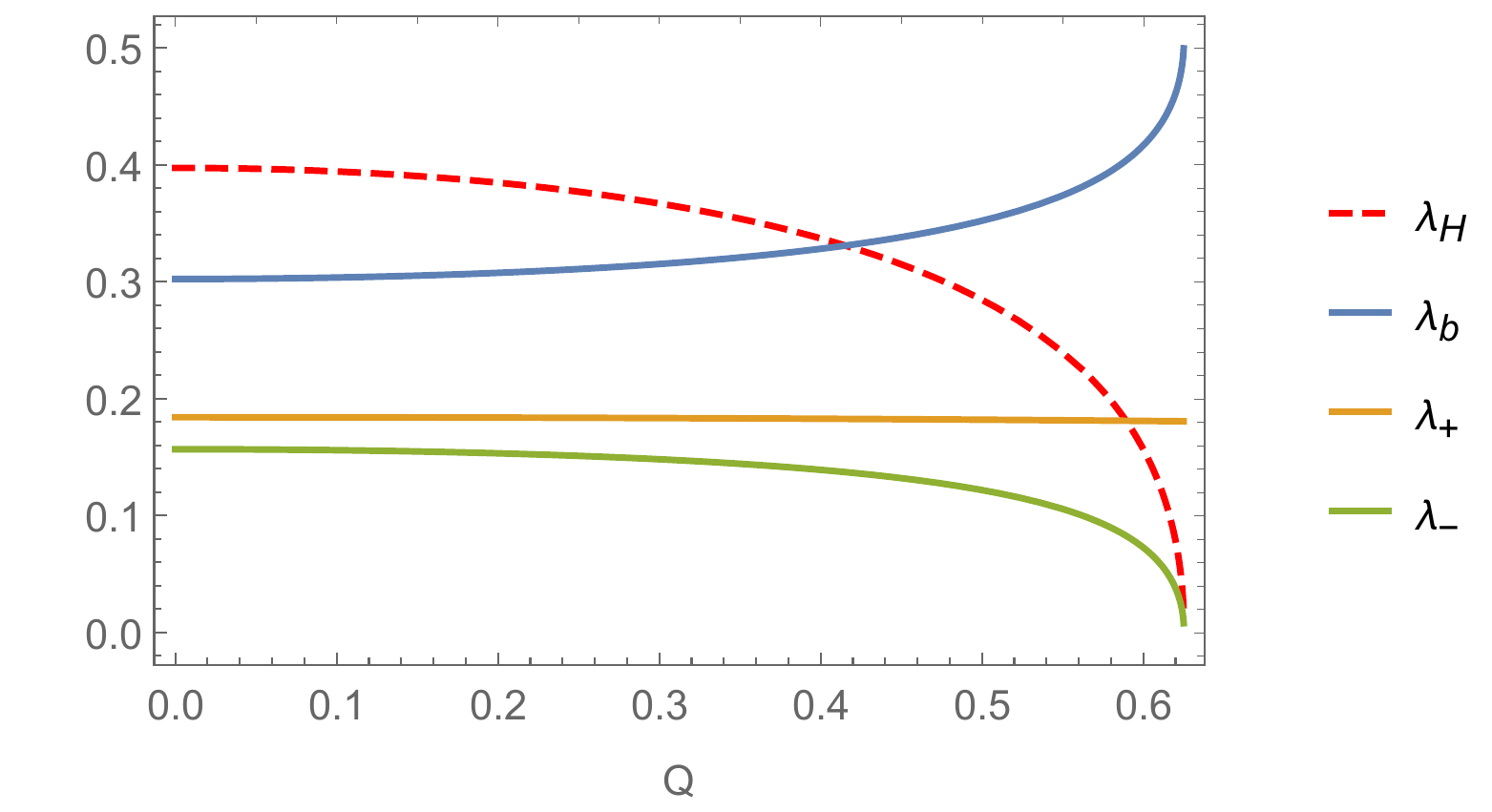}}
\\
\subfloat[]{\includegraphics[scale=0.45]{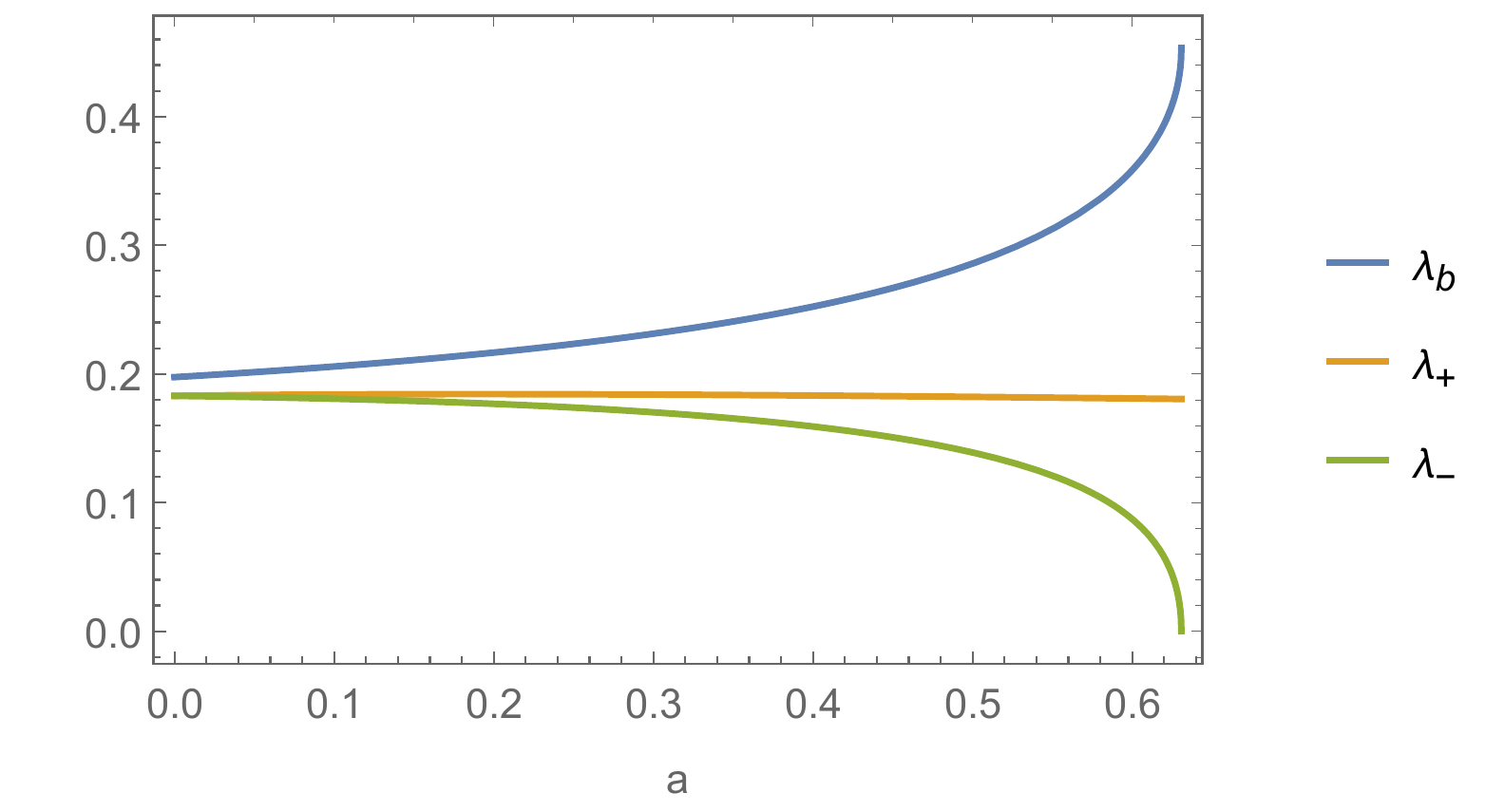}}
\qquad
\subfloat[]{\includegraphics[scale=0.45]{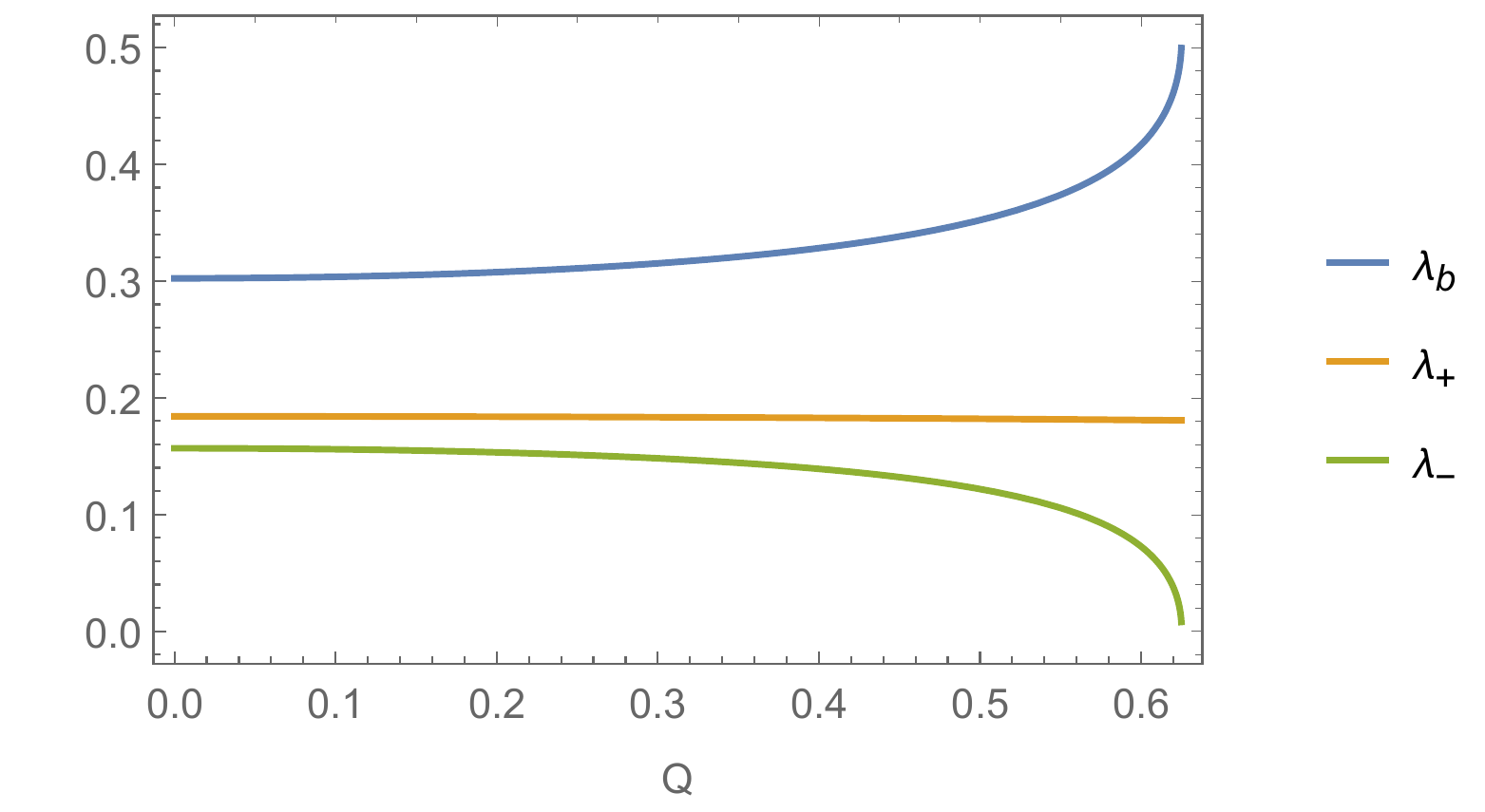}}
\caption{\footnotesize{(a,b) Comparison between the Lyapunov exponents of a BH with $M=1$, $\ell =3$ and its temperature as a function of the rotation parameter $a$ (left, $Q=0.7$) and the BH charge $Q$ (right $a=0.7$). (c,d) The BH temperature has been removed to highlight the hierarchy between $\lambda_b$, $\lambda_+$ and $\lambda_-$.}}
\label{plotslyap}
\end{figure}

In figures \ref{plotslyap} we plot the functions (\ref{lambda}) $\lambda_\pm = \lambda(r_\pm)$, the minimum and maximum values of $\lambda$,
  $\lambda_H$  and 
\be
\lambda_b = {1 \over 2\, b_{{\rm min}} }
\ee
as functions of the angular momentum and charge, where $b_{{\rm min}}=b_c(r_-)$ is the minimal choice of the impact parameter that an impinging massless particle can have without falling into the BH's. We notice that $\lambda \leq \lambda_b$  with the bound saturated for non-rotating and uncharged BH's.

\section{Analytic results for specific BH's}   
 \label{Analytic}

In this section we compute the Lyapunov exponents and critical impact parameters for some simple instances of BH's in $d=4$ and then present some general results for non-rotating BH's in arbitrary dimensions. For simplicity we focus on geodesics with $\dot{\chi}=0$, corresponding 
 to one of the choices $r_c=r_\pm $. 
 In all the examples we compare the results for $\lambda$ with the one for $\lambda_H$ characterising the vanishing rate of the radial velocity at the horizon and given by
 \begin{equation}
\lambda_H=2\pi \,T = {\Delta_r'(r_H)    \over 2(a^2+r_H^2)  } 
\end{equation}

\subsection{Non rotating BH's in AdS}

 We consider first the case of a non-rotating charged BH in AdS, i.e. $a =0$. 
  The metric reads 
\begin{equation}
ds^2=-\frac{{\Delta_r}}{r^2} dt^2+ r^2 \sin^2\theta d\phi^2   + \frac{r^2 dr^2}{{\Delta_r}} + r^2 d\theta^2 \label{adssch}
\end{equation}
with 
\begin{equation}
\begin{aligned}
\Delta_r(r) &=  \frac{r^4}{{{{\ell}}^2}}+r^2-2 M r +Q^2\,,
 \end{aligned}
\end{equation} 

The peculiar feature of the non-rotating case, is that due to full rotational symmetry, the radial velocity of a probe depends only on the total angular momentum $K=bE$, so the critical equations determine both $b$ and the critical radius $r_c$, leaving ${{\zeta}}$ undetermined . 
 The photon-sphere is a round sphere with critical radius given in terms of the BH mass and charge. 
   The  critical equations  reduce to
\begin{equation}
    r_c^4 - b_c^2\, \Delta_r(r_c) =  4 \,   \Delta_r(r_c) -  \, r_c \Delta'_r(r_c) =0
\end{equation}
 that can be solved  for $r_c$ and $b_c$. 
\\
\\
\textbf{Schwarzschild BH in flat space-time}
\\
Setting $Q=0$ and sending $\ell\rightarrow\infty$, one gets a Schwarzchild BH in flat space-time for which 
\be
r_c=3 M \quad, \quad b_c=3\sqrt{3}\, M \quad , \quad \lambda={1\over 2 b_c}
\ee   
 We notice that  
 \be
 \lambda=\lambda_b={1\over 2 b_c} =  {1\over 6\sqrt{3}\, M} <\lambda_H={1\over 4 M} 
 \ee
\\ 
\textbf{Reissner-Norsdtr\"om BH in flat space-time}
\\
Sending $\ell\rightarrow\infty$ with $Q\neq 0$ one finds
\be
b_c = \sqrt{ 2 r_c^3\over r_c-M } \qquad, \qquad 
 \lambda =  {1\over 2b_c} \sqrt{ 2-{3M \over r_c} } 
\ee
with
\be
r_c = \ft12\left(3M+\sqrt{9 M^2-8 Q^2}\right) \quad, \quad 
\ee
and $0<Q\le M$. We notice that $\lambda <  \lambda_b$ in the charged case.  
Comparing with 
\be
\lambda_H={r_H-M\over r_H^2}
\ee
one finds that the bound $\lambda <\lambda_H$ is violated inside the very narrow window $0.99 M< Q\le M$ near extremality, where $\lambda >\lambda_H$.  In particular, in the extremal case $Q=M$, $\lambda_H=0$, while for $\lambda = \lambda_{\rm ext}$ one finds 
\bea
r_c &=&  2\,M \quad, \quad 
b_c = 4\, M \quad, \quad
\lambda_{\rm ext} = { 1 \over 8 \sqrt{2} M}>\lambda_H=0
\eea
\\
\textbf{AdS-Schwarzschild BH}
\\
Setting $Q=0$ with $\ell$ finite, one  finds
\be
r_c=3 M \quad, \quad b_c={ 3\sqrt{3}  \, M\,\ell  \over  \sqrt{ 27 M^2+ \ell^2 }  }    \quad , \quad \lambda={ 1  \over   2 b_c}  \label{bsch}
\ee   
 Solving $\Delta_r(r_H)=0$  for $M$ in favor of $r_H$ , it is easy to check that $\lambda<\lambda_H$ 
 for any $r_H$ and $\ell$.  
\\
\\ 
\textbf{AdS-Reissner-Norsdtr\"om BH}
\\
Keeping $Q\neq 0$ one finds 
\be
  b_c    = \sqrt{2\ell^2 r_c^3 \over 2 r_c^3+\ell^2(r_c-M)}   \qquad, \qquad     \lambda ={1\over  2 b_c} \sqrt{ 2  -{3M \over r_c} } 
  \ee
 with
  \be
  r_c = \ft12 \left(3M+\sqrt{ 9 M^2- 8 Q^2}\right) 
  \ee
Note that $ 2M \leq r_c \leq 3M$, so  
     \be
     \lambda \leq \lambda_b={1\over  2 b_c}
     \ee
      with the bound saturated for $Q=0$. 
 
 \subsection{Rotating BH's}
 
 Expressions for $\lambda$ in the general case of rotating BH's are quite involved so we  
 limit ourselves to special choices of the parameters 
such that analytic formulae can be found. 
\\
\\
\textbf{Large photon-spheres}
\\
In the  limit where the size $\ell$ of Anti-deSitter is much smaller than the radius of the photon-sphere, i.e. $a,\ell<<M,Q, r_c$  one finds
\be
b_\pm = \mp \,\zeta_\pm  \approx  \ell    \qquad , \qquad   
 r_\pm \approx  {3\, M +\sqrt{9M^2 - 8Q^2(1\mp{ a\over \ell})^2 }   \over 2 (1\mp{ a\over \ell})^2}       \\
\ee
 For the Lyapunov exponents one finds
 \begin{equation}
\begin{aligned}
\lambda_\pm &\approx    {\sqrt{3 M\, r_\pm-4 Q^2} \over  2 \ell\, r_\pm\,   (1\mp { a\over \ell})}   < {\sqrt{3 M\, r_\pm-2 Q^2} \over  2 \ell\, r_\pm\,   (1\mp { a\over \ell})} ={1\over 2\ell} =\lambda_b
\end{aligned}
\end{equation}
as expected.
%
%
\\
\\
\textbf{Extremal asymptotically flat Kerr-Newman BH}
\\
The extremal Kerr-Newman  BH in flat space-time ($\ell=\infty$) is obtained by taking 
\be
M=r_H \quad, \quad Q^2=r_H^2-a^2
\ee
leading to 
\be
\Delta_r=(r-r_H)^2
\ee
One finds 
\bea
r_- &=&  2(r_H - a) \quad ,\quad    b_-= {{\zeta}}_-  =2\, (r_H-a) \nn\\
r_+ &=&  2(r_H + a) \quad ,\quad    b_+= -{{\zeta}}_+  =2\, (r_H+a) 
\eea
   Here we assume that  $a<r_H/2$, in order to ensure that $r_-$ is outside the BH horizon. For the Lyapunov exponents one finds
\be
\lambda_\pm = {1\over 2 (4 r_H-3 a) }  \left({r_H-2a \over 2r_H-2a }\right)^{3\over 2}  
\ee
 We notice that $\lambda>\lambda_H=0$, so the bound is violated as claimed. 
 
\subsection{Spherical symmetric BH's in higher dimensions}
  
We now extend our analysis and consider critical null geodesics around spherically symmetric BH's in $d$-dimensions 
  with AdS or flat asymptotics.  The line element of an asymptotically $AdS_d$ spherically symmetric BH reads
\be 
ds^2 = -f(r)dt^2 + f(r)^{-1}dr^2 + r^2 ds^2_{S^{d-2}}
\ee   
where
\be 
f(r) = 1 - \frac{2M}{r^{d-3}}+ \frac{Q^2}{r^{2(d-3)}}+\frac{r^2}{\ell^2}
\ee
 and $M$, $Q$, $\ell$ describing the mass, charges and AdS radius of the asymptotic geometry. 
The Hamiltonian null condition reads
\be
{\cal H}=f(r)\, P_r(r)^2  -{E^2\over f(r)}+{K^2\over r^2}=0
\ee
with $K$ denoting the total angular momentum. 
  The radial velocity reads
\be 
\frac{dr}{dt} = \frac{f(r)^2\,P_r}{E}    = { f(r) \over r} \sqrt{\mathcal{R}(r)}
\ee
with  $b= K/E$ the impact parameter and
\be 
\mathcal{R}(r) = r^2 -b^2\, f(r)  
\ee
Near the horizon one finds the radial velocity vanishing as
   \be
 {dr\over dt}\approx -2 \lambda_H (r-r_H)
 \ee
 with
 \be
\lambda_H =\frac{f'(r_H)}{2 r_H}\sqrt{\mathcal{R}(r_H) }  =  \frac{f'(r_H)}{2 }  =   2\pi T
\ee  
given in terms of the BH temperature $T$. 
The BH  photon-sphere is defined by the critical equations $\mathcal{R}(r_c)=\mathcal{R}'(r_c)=0$  leading to
\be 
\begin{aligned}
 b_c^2={r_c^2 \over f(r_c) }   
\end{aligned}
\ee
with $r_c$ solving 
\be
r_c\, f'(r_c)  =2 f(r_c) 
\ee
Near the critical radius one finds
 \be
 {dr\over dt}\approx -2 \lambda (r-r_c)
 \ee
 with
 \be
\lambda=   \frac{f(r_c)}{2 r_c}\sqrt{\mathcal{R}^{''}(r_c) \over 2} =  \frac{\sqrt{d-3}}{2\,b_c}\sqrt{ 1-(d-2) {Q^2\over r_c^{2(d-3)} } }
\ee
Note that $\lambda$ is bounded from above
  \be
  \lambda \leq  \lambda_b=\frac{\sqrt{d-3}}{2\,b_c}
  \ee
  with the bound saturated by the uncharged BH's. For $Q=0$ one finds 
  \be
  \lambda=  \frac{\sqrt{d-3}}{2\,b_c} 
  \ee
  with
 \be 
b_c ={r_c \over \sqrt{f(r_c)} }   \qquad , \qquad r_c^{d-3} =  2M  \left(\frac{d-1}{2}\right) 
\ee
Quite remarkably $r_c$ does not depend on $\ell$ but $b_c$ does.
 
\section{Fuzzball geometries}
 \label{Fuzzballs}
 
 In this section we study critical and nearly critical null geodesics in a class of three-charge microstate geometries of type $(1,0,n)$ first introduced in \cite{Bena:2016ypk, Bena:2017xbt}. 
 We consider both the asymptotically flat and the near-horizon limit exhibiting pretty different behaviour. We focus on scattering along the $\vartheta=0$ direction,  where the geodesic equations have been shown to be integrable in this class of microstate geometries. We will also set $Q_1=Q_5=L^2$ for notational simplicity. The geometries are 
characterized by an asymptotically $AdS_3\times S^3$ metric  that after a conformal rescaling\footnote{We recall that conformal rescalings are irrelevant for the study of null geodesics.}  can be written as
 \be
 \begin{aligned}
ds_6^2=   -2  \left(dv+\beta \right)\left( du  +  \gamma  \right) +  Z^2 \, ds_4^2   \,.
\end{aligned}
\ee
with 
 \be
\begin{aligned}
\label{metricQ1Q5Qpds4}
ds_4^2=\left(\rho^2{+}a^2\cos^2 {{\vartheta}}\right)\left(\frac{d\rho^2}{\rho^2{+}a^2}+d{\vartheta}^2\right)+\left(\rho^2+a^2\right)\sin^2 {{\vartheta}} \,  d\varphi^2 + \rho^2 \cos^2 {\vartheta}\, d\psi^2 \, .
\end{aligned}
\ee
the flat space metric on $\mathbb{R}^4$ in oblate spheroidal coordinates. For $\vartheta=0$, the various functions entering in the metric take the form
 \bea
Z & =&1+ \frac{ L^2}{\rho ^2+a^2  } \quad, \quad \beta =-\frac{ a^2 R \,  }{  \rho^2+a^2 } \, d\psi   \quad, \quad
\gamma=    \frac{ a^2 R \,    (1-{\cal F}_n)  }{  \rho^2+a^2 }    d\psi  +\mathcal{F}_n\, dv   
\eea
and\footnote{In the notation of \cite{Bianchi:2018kzy} $\epsilon_1=\epsilon_4^2=2 a^2(\nu-1)$. }
\be
\label{FepsilonDef}
\begin{aligned}
  \mathcal{F}_n(\rho)  &= \left(1- \nu  \right) 
  \left[ 1- \left({\rho^2 \over \rho^2+a^2}\right)^n  \right]      \quad, \quad   \nu\equiv \frac{L^4}{ 2 a^2\,R^2}   
\end{aligned}
\ee
  Motion along the $\vartheta=0$ direction requires $P_\vartheta=P_\varphi=0$, so the Hamiltonian reduces to
  \be
\label{nulllambda}
\begin{aligned}
 & \mathcal{H}  =  {P_{\rho }^2 \over 2\, Z^2} - P_u(P_v-{\cal F}_n\, P_u)+{\left[ P_\psi(\rho^2+a^2)+a^2 \, R\, (P_v- P_u) \right]^2 \over 2 \,Z^2\,\rho^2\, (\rho^2+a^2)^2}
\end{aligned}
\ee
We set
\bea
\label{EPyDef}
P_u= {E+P_y\over \sqrt{2}}   \quad, \quad   P_v = {E-P_y\over\sqrt{2}}
\eea
and focus on the case $P_y=0$. For this choice the Hamiltonian reduces to
 \be
 {\cal H}= {\rho^2\,P_{\rho }^2 +P_\psi^2 \over 2\, Z^2\, \rho^2} - {E^2\over 2}(1-{\cal F}_n) 
 \ee
 and the velocities w.r.t. the radial variable are given by
\be
\begin{aligned}
\frac{dt}{d\rho} =-\frac{  {\partial\mathcal{H}}/{\partial E}}{{\partial\mathcal{H}}/{\partial P_\rho} } = {Z(\rho)^2  E \left[ 1-{\cal F}_n(\rho)\right] \over P_\rho(\rho)}  \quad ,\quad 
\frac{d\psi}{d\rho}  =\frac{{\partial\mathcal{H}}/{\partial P_\psi}}{{\partial\mathcal{H}}/{\partial P_\rho} } = -{  P_\psi 
\over P_\rho(\rho)\, \rho^2  } 
\end{aligned}
\ee
 with
 \be
  {P_{\rho}(\rho) \over E}   =  \sqrt{ {\cal R}(\rho) }=\sqrt{ Z(\rho)^2 \left[1-{\cal F}_n(\rho) \right]-{  b^2    \over \rho^2 }}
 \ee
 and $b=P_\psi/E$. We notice that positivity of  ${\cal R}(\rho)$ at infinity requires
 \be
 b^2 < 2 L^2
 \ee
  The critical equations 
\be
{\cal R}(\rho_c)={\cal R}'(\rho_c)=0
\ee
can be solved for $b_c$ and $\rho_c$.  
 Near the critical radius one finds
 \be
 {d\rho\over dt}\approx -2 \lambda (\rho-\rho_c)
 \ee
 with
 \be
\lambda= {\sqrt{ {\cal R}''(\rho_c) / 2 } \over 2 Z^2(\rho_c) (1-{\cal F}_n(\rho_c) )}
 \ee

\subsection{The BH geometry}

The BH geometry corresponding to the above micro-states is obtained by sending $a\to 0$, $\nu\to \infty$ and $n \to 0$ keeping finite the product
\be
L_p^2=n\, \nu \, a^2    
\ee
leading to
\be
  {\cal F}_n =-{L_p^2\over \rho^2}    \qquad ,\qquad     Z =1+ \frac{ L^2}{\rho ^2   }
\ee
The radial function ${\cal R}(\rho)$ becomes
\be
{\cal R}(\rho)=1+{3L^2-b^2\over \rho^2}+{3L^4\over \rho^4}+{L^6\over \rho^6}
\ee
The critical radius and impact parameters read
\be
\rho_c =\sqrt{2} \, L  \qquad ,\qquad b_c={3\sqrt{3}\over 2}L
\ee
leading to the Lyapunov exponent
\be
\lambda_{BH}= {2\over 9 L}  \label{lambdabh}
\ee
 We notice (fig. \ref{flambdafuzz}) that $\lambda_{BH}<1/(\sqrt{2} b_c)$. 
 The asymptotically AdS solution can be analysed by simply dropping the 1 in $Z$. It is easy to see that no solutions to the critical equations are found in this case.  

 \subsection{The asymptotically flat fuzzball }
\begin{figure}
\centering
\subfloat[]{\includegraphics[scale=0.55]{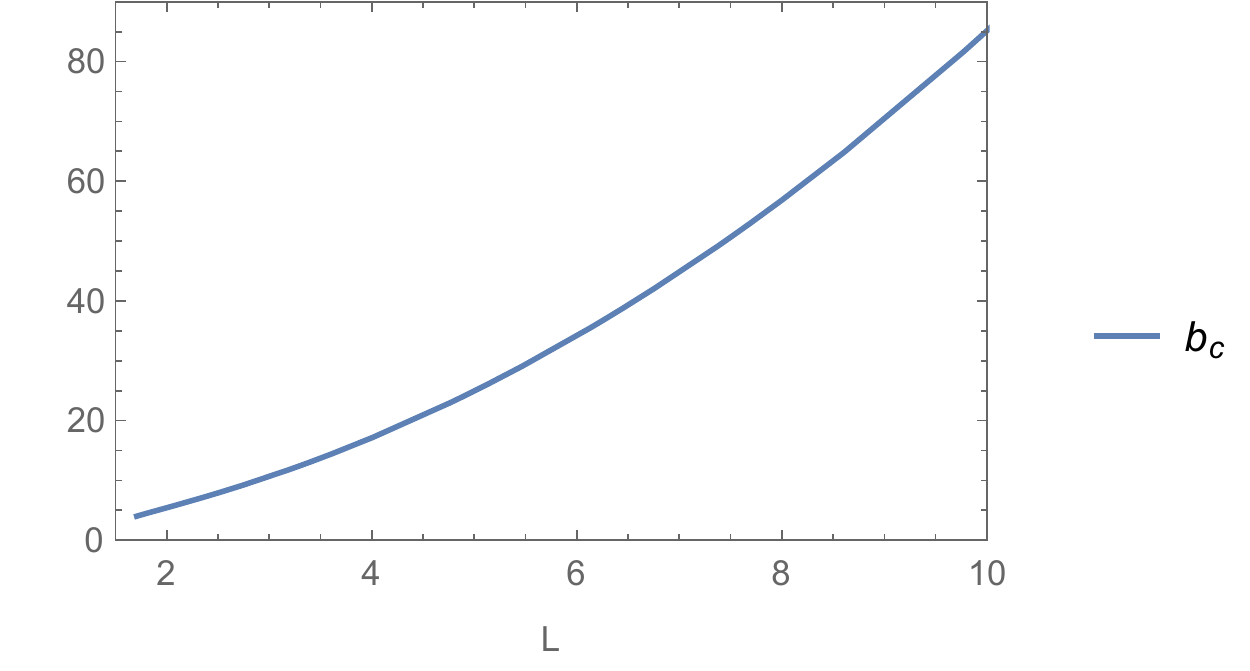}}
\,
\subfloat[]{\includegraphics[scale=0.55]{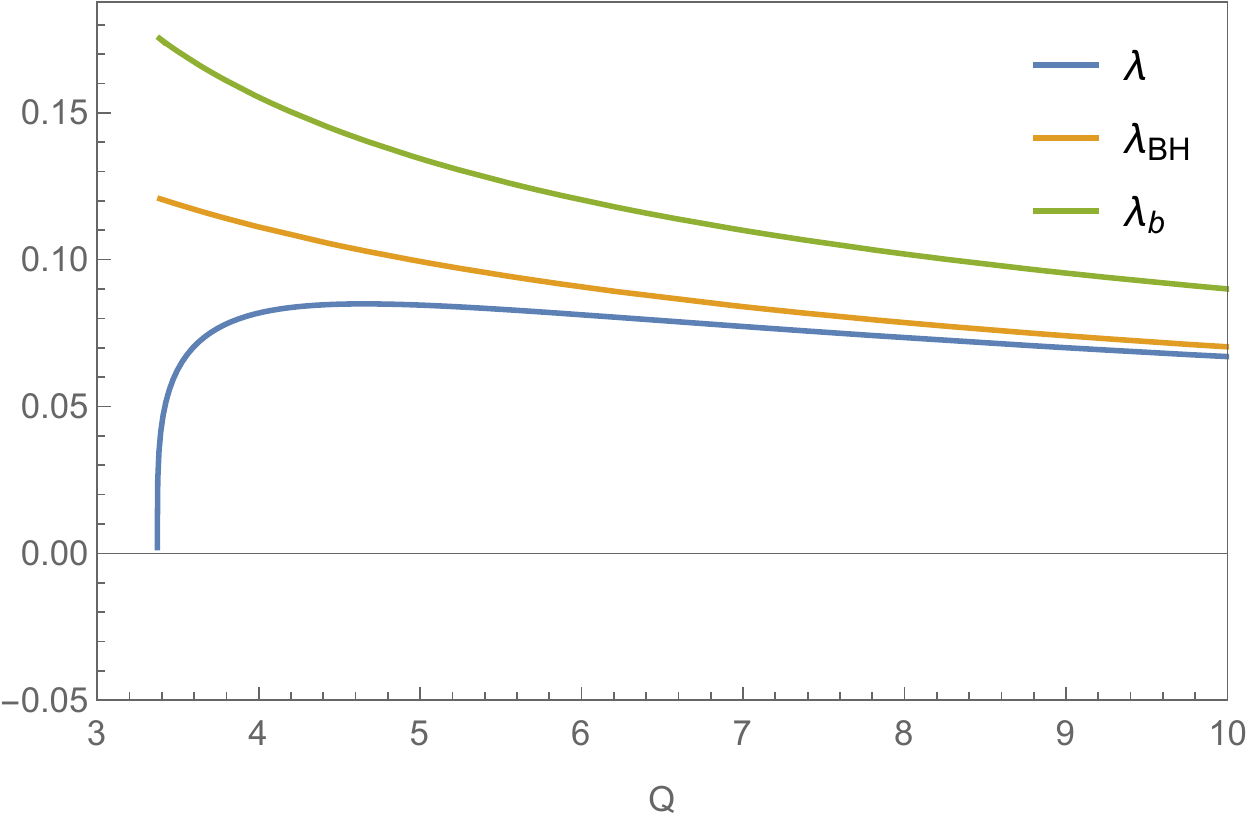}}
\caption{\footnotesize{Critical impact parameter $b_c$ (a) and Lyapunov exponent (b) as a function of the fuzzball charge $L$, for $a =1$, and $n=1$.}}
\label{flambdafuzz}
\end{figure}
  
\begin{figure}[t]
\centering
\subfloat[]{\includegraphics[scale=0.5]{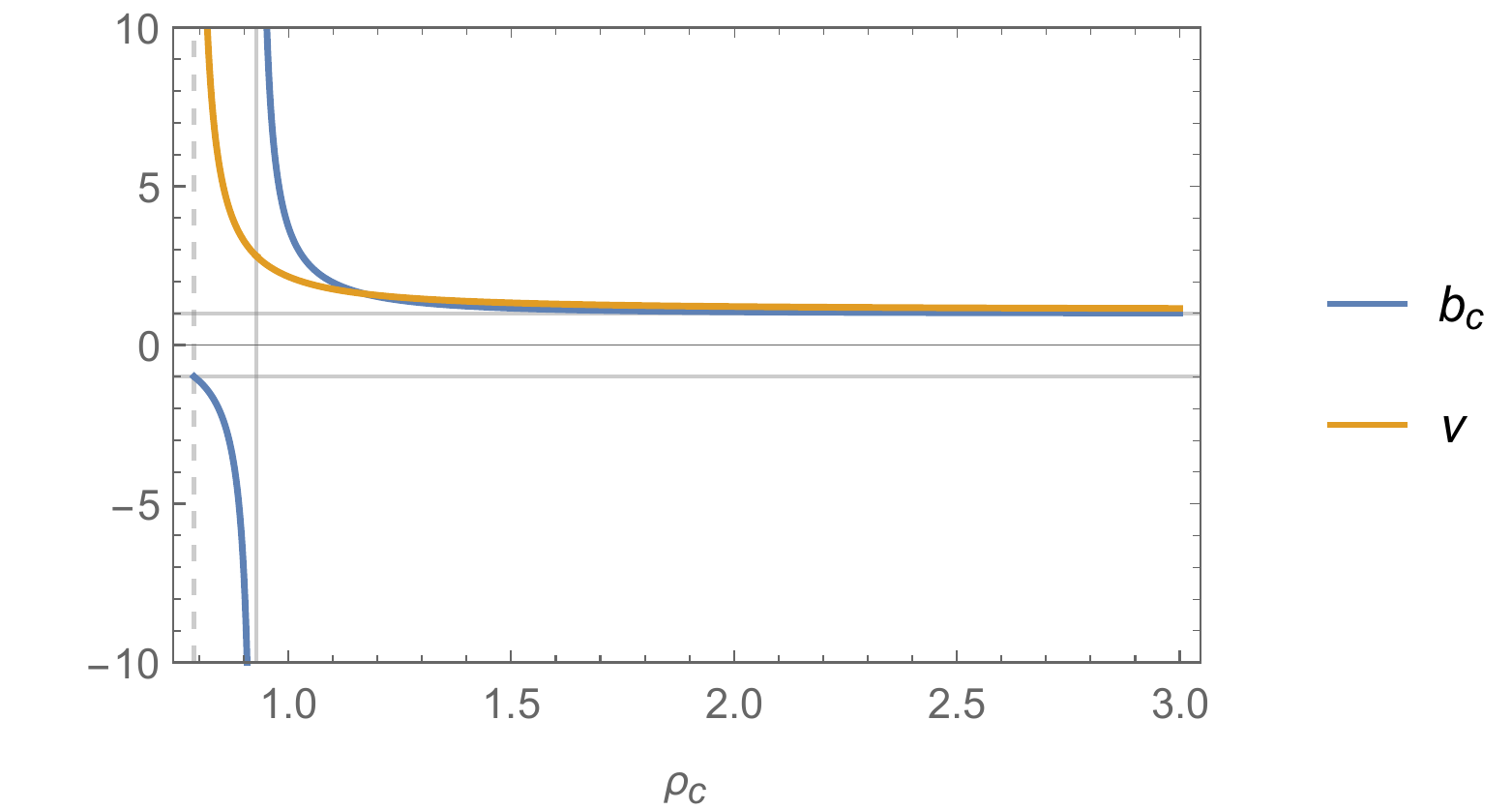}}
\subfloat[]{\includegraphics[scale=0.5]{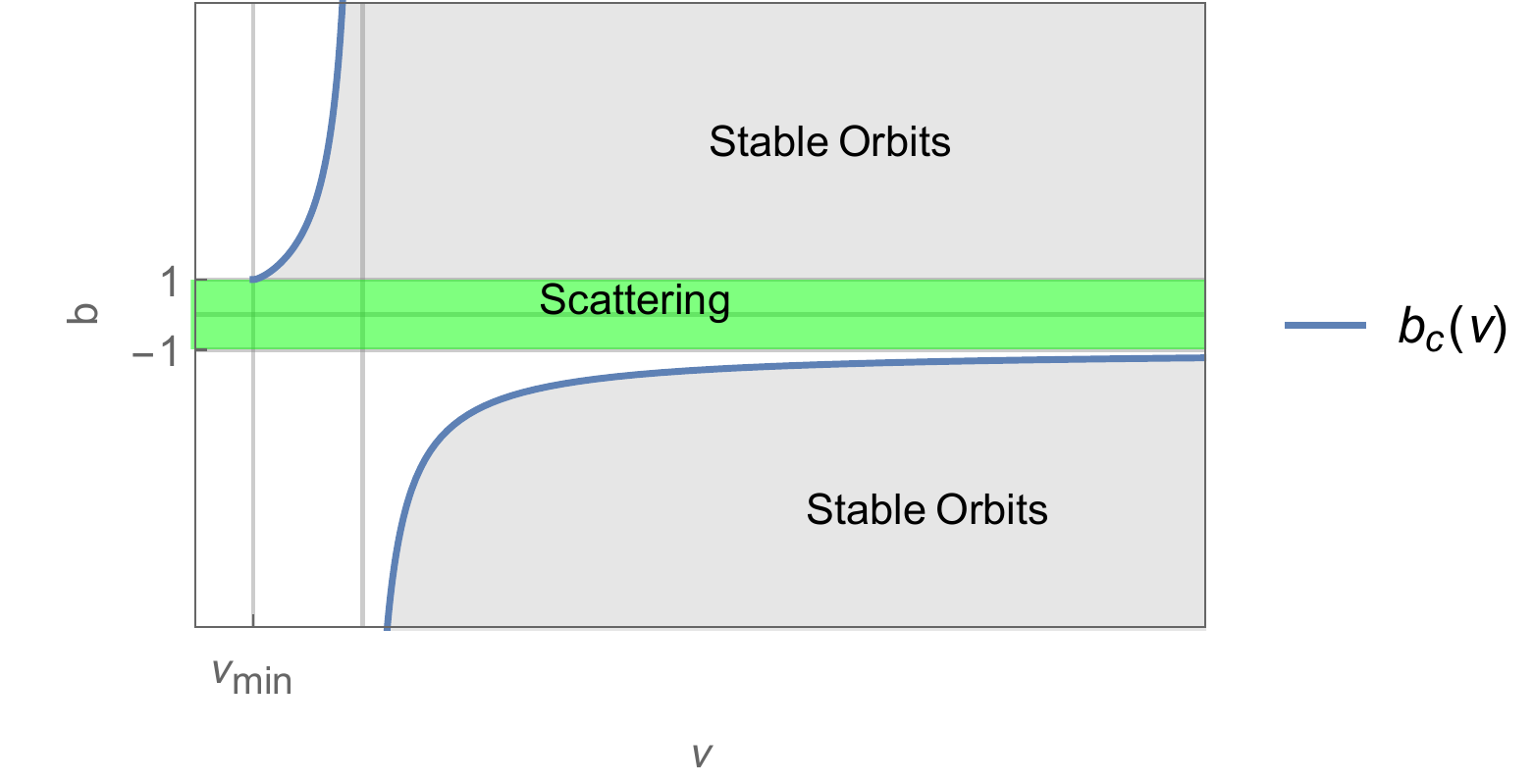}}
\caption{\footnotesize{(a) Critical impact parameter $b_c$ and geometric parameter $\nu$ allowing  for circular phton orbits as a function of the critical radius $\rho_c$, $a$ is set to 1. (b) Parameter space for null geodesics with zero total angular momentum.}}
\label{byVSvareps}
\end{figure}

    In the case of the asymptotically flat fuzzball the critical equations  can be written in the form
 \be
 \begin{aligned}
 & b_c(\rho_c) =  \rho_c\,  Z(\rho_c)   \sqrt{1-{\cal F}_n(\rho_c,\nu)} \\
& 2 \left[1-{\cal F}_n(\rho_c,\nu)\right](\rho_c\, Z'(\rho_c)+Z(\rho_c))-Z(\rho_c) \, \rho\, {\cal F}'_n(\rho_c,\nu)=0
 \end{aligned}
 \ee
 The second equation is linear in $\nu$ and can be solved for $\nu=\nu(\rho_c)$ that then can be plugged into the first equation to determine $ b_c(\rho_c)$. 
 For instance for $n=1$ one finds
 \be
  \begin{aligned}
   b_c(\rho_c) &=  {\rho_c^2    \over   (\rho_c^2+a^2) }  \sqrt{ (\rho_c^2+a^2+L^2)^3 \over  \rho_c^2(2L^2-a^2)-a^4-a^2 L^2  } \\
 \nu(\rho_c) &=  {\rho_c^2 [\rho_c^4+\rho_c^2(3a^2-L^2)+2 a^4+2 a^2 L^2 ] \over a^2\left[\rho_c^2(2L^2-a^2)-a^4-a^2 L^2 \right] }\\
  \end{aligned}
 \ee
 Near the critical radius one finds
 \be
 {d\rho\over dt}\approx -2 \lambda (\rho-\rho_c)
 \ee
 with
 \be
\lambda= {\sqrt{ {\cal R}''(\rho_c) / 2 } \over 2 Z^2(\rho_c) (1-{\cal F}_n(\rho_c) )}
 \ee
 The functions $b_c(\rho_c)$,  $\nu(\rho_c)$, $\lambda(\rho_c)$ can be used to plot parametrically $b_c(\nu)$, $\lambda(\nu)$.
 The plots are displayed in figure \ref{byVSvareps}  for a fuzzball solution with $n=1$ and $L=5$. 
 Plot \ref{flambdafuzz} shows that $\lambda$ is below the bound (in $d=5$)
 \be
\lambda< \lambda_b={1 \over \sqrt{2} b_c} 
 \ee
 We notice also that  $\lambda$ of the fuzzball is smaller than the Lyapunov exponent $\lambda_{BH}$ (\ref{lambdabh}) associated to the BH of the same charge. 
\subsection{Asymptotically AdS geometry}
The asymptotically AdS case is obtained by dropping the 1 term in $Z$.\footnote{The metric is asymptotically conformally AdS with radius $\ell = \sqrt{L_1 L_5} = L$.} The positivity of $P_{\rho }^2$ at infinity requires $P_\psi=0$ so only geodesics moving entirely inside $AdS_3$ can arrive from the AdS boundary\footnote{ Photon orbits with non-trivial angular momentum $P_\psi$ exists and were studied in \cite{Bianchi:2018kzy}.}. 
  
Starting from infinity, geodesics evolve untill they reach the turning point $\rho_*$, i.e. a zero of $P_\rho(\rho)$, where the radial velocity vanishes.
Zeroes of $P_\rho(\rho)$ coincide with the ones of the function
 \be
{{\cal R}_n}(\rho) = {\,({\rho^2{+}a^2})^2 \, P_\rho^2 (\rho)\over 2\nu\, a^2 \,E^2\, R^2}  = 1-b^2\left( 1+{a^2\over \rho^2\, \nu}\right) -{\cal F}_n(\rho)(1+b)^2 \label{p3c}
\ee
where we introduced the impact parameter
\bea
\label{bDef}
b = {P_y\over E}
\eea
 
The critical equations 
\be
{\cal R}(\rho)={\cal R}'(\rho)=0
\ee
can be written in the form
\be
\begin{aligned}
 & a^2   \left[2-2{\cal F}_n(\rho_c)-\rho_c{\cal F}'_n(\rho_c)\right]^2 -  8 \nu \rho_c^3\, {\cal F}'_n(\rho_c) =0\\
& b_c =  {2-2{\cal F}_n(\rho_c)-\rho_c{\cal F}'_n(\rho_c)\over 2+2{\cal F}_n(\rho_c)+\rho_c\,{\cal F}'_n(\rho_c)}  \label{critbfuz} 
\end{aligned}
\ee
The first equation is quadratic in $\nu$, since $\mathcal{F}_n$ is linear in $\nu$, and we can solve it as a function of $\rho_c$. The second equation determines $b_c(\rho_c)$
once $\nu$ is replaced by $\nu(\rho_c)$ inside $\mathcal{F}_n$. One finds non trivial solutions for $n\geq 2$.  The explicit form of the solutions are not particularly illuminating, so we will not display it here.  

In figure \ref{byVSvareps} we show the curves $b_c(\rho_c)$ and  $\nu(\rho_c)$ for $n=2$. First, we notice that there is no solution in the scattering region $|b|<1$ where motion is allowed from infinity, so critical geodesics exists but they cannot be reached starting from infinity. Second we observe that for $b\approx \pm 1$, 
equations (\ref{critbfuz}) admit critical solutions for
\bea
b_c &\approx & 1   \qquad , \qquad  \nu \approx \nu_{\rm min}=\ft12+\ft12\sqrt{n+1\over n} \qquad, \qquad \rho_c>>a\nn\\
b_c &\approx & -1   \qquad , \qquad  \nu >>1   
\eea
  For $\nu<\nu_{\rm min}$ no critical geodesics exists. 
     We find that geodesics fall into two different categories: scattered geodesics impinging from infinity and finding at most one turning point (green region), and stable orbits
  confined in the interior of AdS (grey regions) bouncing between two turning points.
   
\begin{figure}[t]
\centering
\subfloat[]{\includegraphics[scale=0.5]{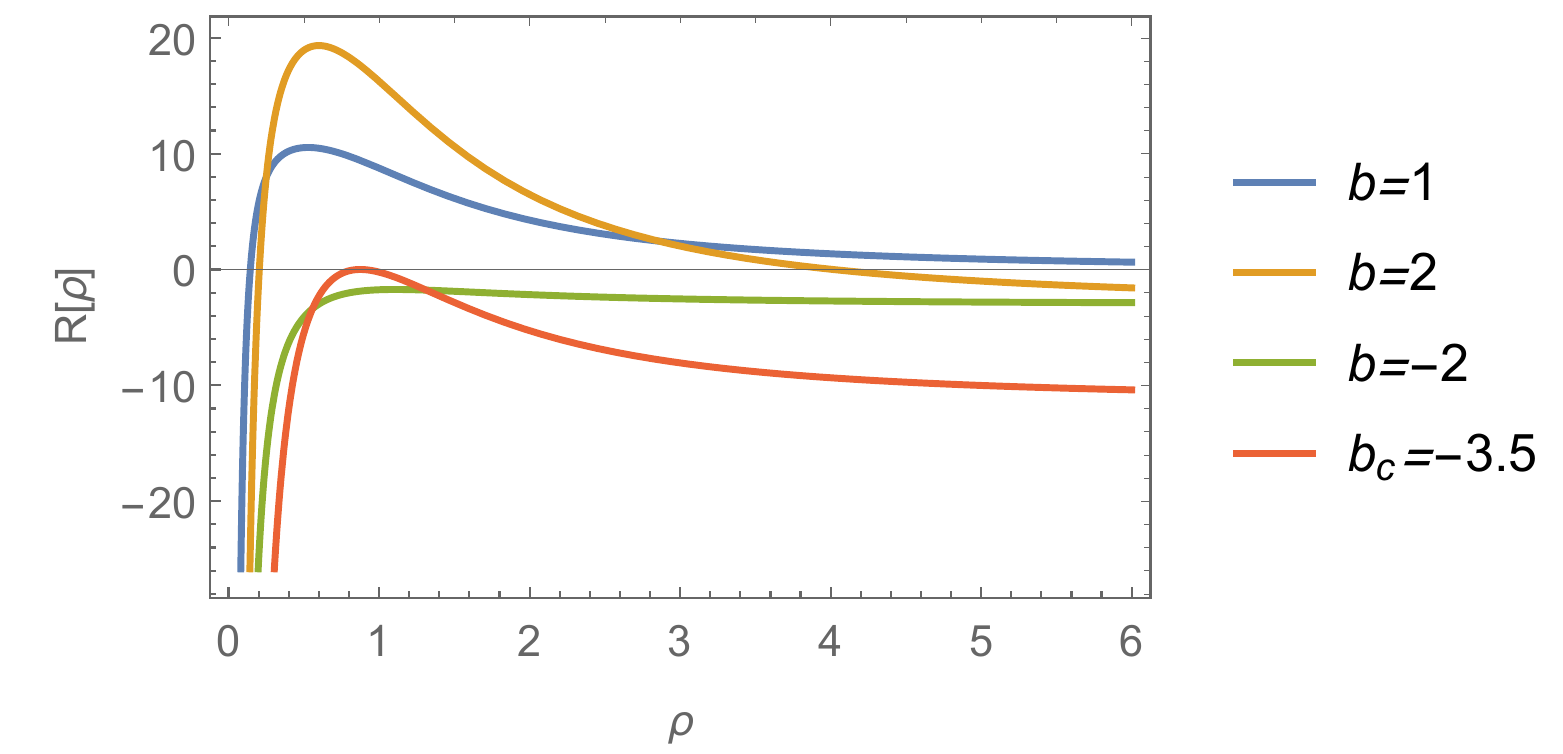}}
\subfloat[]{\includegraphics[scale=0.5]{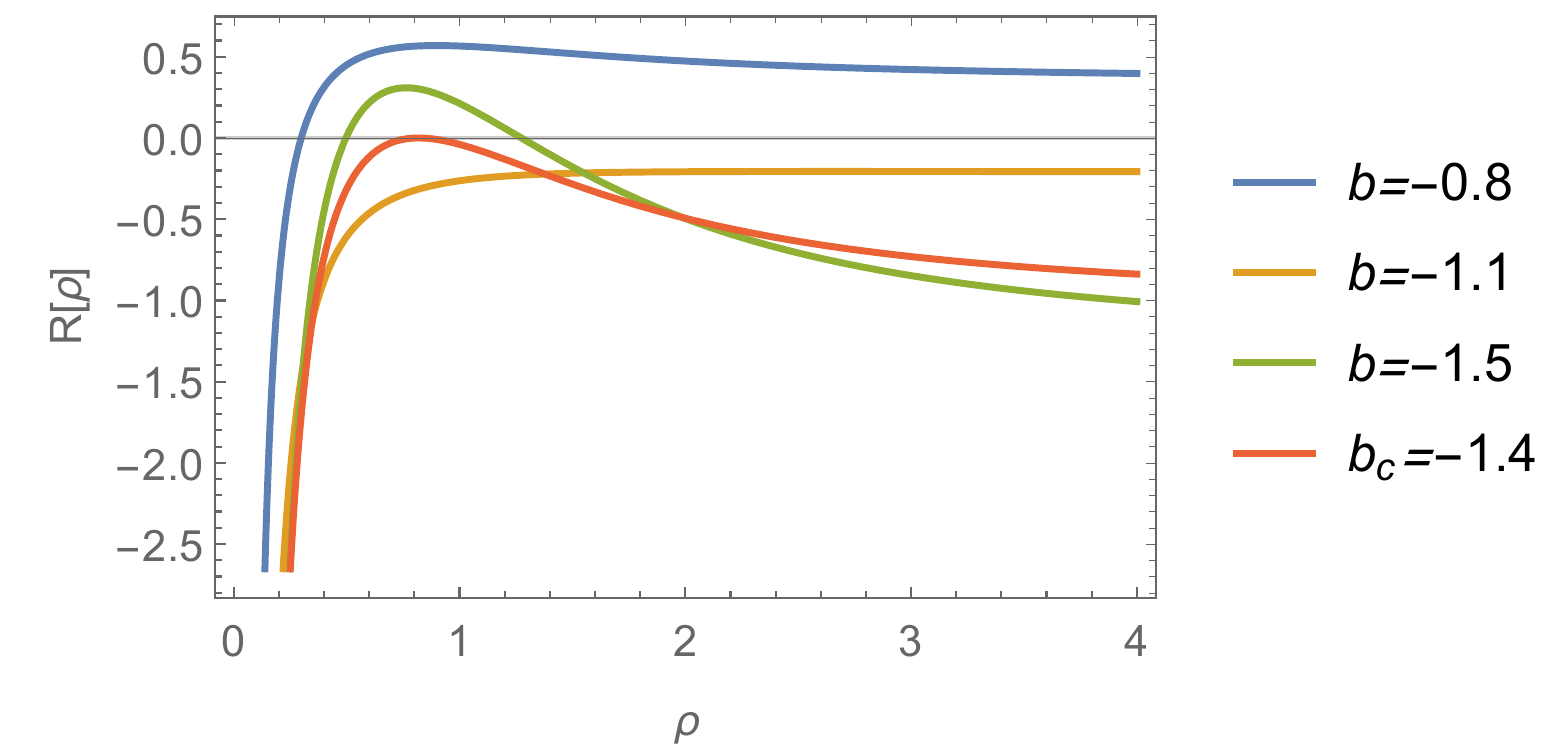}}
\caption{\footnotesize{Radial function $\mathcal{R}(\rho)$ as a function of $b$ and $\rho$ for a) $\nu=4$ and b) $\nu=10$.}}
\label{fuzzregions}
\end{figure}

  This behavior is shown in figure \ref{fuzzregions}, where we draw the radial function $\mathcal{R}(\rho)$ for different values of the impact parameter $b$. For values of $b$ in the green region the particle comes from infinity and reaches a turning point when $\mathcal{R}(\rho_T)=0$ (\textit{e.g.} blue curves in plot (a) and (b)), for values of $b$ in the grey region the particle bounces back and forth between two turning points (orange curve in (a), green curve in (b)).
  
   We conclude that the photon-sphere of fuzzballs in the class $(1,0,n)$ under study cannot be reached by geodesics scattered from the AdS boundary at infinity.

  \section{Discussion and outlook} 
 \label{Conclusions}

 We have studied geodesics for massless neutral probes around the photon-sphere, discriminating between the scattering and absorption phases of asymptotically AdS Kerr-Newman BH and fuzzball geometries. We have found that geodesics near the photon-sphere exhibit always a chaotic behaviour characterised by  an exponential growth of the angular dispersion of nearby geodesics.  
  We have computed the critical impact parameters and the Lyapunov exponent $\lambda$ governing the exponential growth as functions of the radius of the limiting photon orbits. We have related the Lyapunov exponent $\lambda$ to the ratio between the vanishing radial velocity and the distance from the photon-sphere
 \be
  \dot r \approx  2 \lambda\, (r-r_c)  \quad {\rm as} \quad r\approx r_c
  \ee  The coefficient $\lambda$ depends on the parameters characterizing the BH: mass, charges and angular momentum.  Geodesics with impact parameters below the critical values fall into the horizon with vanishing radial velocities again proportional to the distance from the horizon, with the ratio $\lambda_H$  given solely in terms of the BH temperature $\lambda_H=2\pi T$.
   
    The critical exponent $\lambda_H$ has been recently related to the Lyapunov exponent computing
   the exponential growth of out-of-time ordered correlators in the quantum mechanic systems describing holographically the near horizon BH geometry  \cite{Maldacena:2015waa}.  A quantum system dual to a BH is expected to be maximally chaotic, so $\lambda_H$ has been proposed as an upper bound on quantum chaos a thermal theory at temperature $T$ can develop. The BH photon-sphere is far away from the horizon, so chaos in the photon-sphere is not related 
   to that in the quantum mechanics in any obvious way.  Still it is interesting to observe that $\lambda$
   is typically below the advocated upper bound $\lambda_H$,  but the bound is badly violated  inside a very narrow window near extremality, where the BH temperature $T_H$ is very small or vanishing and the photon-sphere coalesces with the BH horizon. Due to the obvious limitations of our classical analysis, we cannot exclude that the violation of the bound  could be attributed to quantum effects at low temperatures.  Another possibility is that the bound $\lambda<\lambda_H$ should be revised/reconsidered in near extreme conditions even after including quantum or stringy corrections\footnote{in particular, the `quantisation' imposed by consistent angular motion, discussed in the Appendix, at the quantum level may lead to a `quantisation' of $r_c$. We thank G. Festuccia for suggesting this possibility.}
   
   In arbitrary space-time dimension $d$, we have found instead that  $\lambda$ is bounded by  
   \be
   \lambda<\lambda_b={\sqrt{d-3} \over 2 b_c}   
   \ee
   with $b_c$ the minimal impact parameter that a massless particle can travel without falling into the BH.   
      
Finally we have considered null geodesics in fuzzball geometries of the so called $(1,0,n)$ class \cite{Bena:2016ypk, Bena:2017xbt}, with asymptotically flat or $AdS^3\times S^3$ geometries. We have shown that critical photon orbits   exists in both cases, but they can be reached from infinity only in the asymptotically flat case. 
We have also found  that the critical exponent $\lambda$ of the fuzzball is typically smaller than the one of the corresponding BH. This suggests that fuzzball can be distinguished from BH's by the signature of the quasi-normal modes characterising the response of the geometry to small perturbations, that are known to be dominated at late times by a basic mode with time scale given by the inverse of $\lambda$ \cite{Cardoso:2008bp}. One should however keep in mind that the geometries we have considered correspond to a very special class of somewhat  atypical BH micro-states. Further investigation is necessary before reaching a convincing conclusion.\footnote{We would like to thank the referee for raising this point.}. 
   
Since the scattering processes under study here can be related to out-of-time-order correlators in the boundary theory (at `radial' infinity), one expects that the chaotic behaviour of geodesics reflects into an exponential growth of the correlators  in the boundary theory (at `radial' infinity) with Lyapunov exponent $\lambda$. High energy scattering in AdS has been also related to correlators involving the insertion of two massless (light-like separated) states (the incoming and outgoing rays)  and two heavy states (the BH) in the boundary theory \cite{Kulaxizi:2018dxo}. 
  The time delay and deflection angles experienced by the geodesics measure the derivatives of the (eikonal) phase shift (encoding the relevant part of the correlator) with respect to the energy and angular momentum and determine the location of its poles in coordinate space  \cite{Karlsson:2019qfi}.
  One may expect that the results here may provide a detailed prediction about the location of poles in the phase shift of the dual correlators at strong coupling.   
   A complementary picture of the exponential fall off of thermal correlators at large imaginary frequency was obtained in \cite{Festuccia:2005pi} from the study of space-like geodesics  probing regions inside the BH horizon. It would be interesting to explore similar connections in the vicinity of the photon-sphere.

\begin{appendix}
 
 \section{Integrating the angular motion}
 \label{Angular}
 
A different source of chaotic, or more precisely quasi-periodic, behaviour lies in the angular dependence of the metric (\ref{metric}). In order to address the existence of such behaviour we investigate the relation between the two angular variables $\phi$ and $\theta$.

Being interested in critical geodesics, we know that the angular variable $\theta$ is constrained by the values $\pm \chi_{\rm p}$   as shown in section (\ref{ThetaMotion}).

For the non rotating BH's (\textit{i.e.} $a=0$) one gets
 \bea
 {{\varTheta}}(\chi)=  -b^2 \chi^2+ b^2-{{\zeta}}^2    =-b^2(\chi^2 -\chi_{\rm p}^2)    
 \eea
with
\begin{equation}
 \chi^2_{\rm p}=1-{{{\zeta}}^2\over b^2} 
 \end{equation}
The motion in the $\phi$-direction is given by
\begin{equation}
\Delta \phi = \int_{-\chi_{\rm p}}^{\chi_{\rm p}}\frac{\zeta\,b^{-1} d\chi}{(1-\chi^2)\sqrt{(\chi + \chi_{\rm p})(\chi -\chi_{\rm p})}} = \pi + 2\pi n
\end{equation}

As one would expect for critical geodesics in static BH's, while $\cos\theta$ sweeps the interval $[-\chi_{\rm p},\chi_{\rm p}]$ back and forth, the angle $\phi$ spans its whole $2\pi$ domain (an integer number of times) and the particle trajectory is a full circle.

Rotating BH's entail a way more complicated relation between $\theta$ and $\phi$
\begin{equation}
\Delta\phi={\rm i} \int_{-\chi_{\rm p}}^{\chi_{\rm p}} \frac{d\chi}{\sqrt{1-\frac{b^2}{\ell^2}}}\left(\frac{1+\frac{\zeta}{a}}{1-\chi^2}-\frac{1+\frac{a\,\zeta}{\ell^2}}{1-\frac{a^2}{\ell^2}\chi^2}  \right)
\frac{1}{\sqrt{(\chi^2-\chi_{\rm p}^2)(\chi^2-\chi_{\rm m}^2)}}
\end{equation}
 The integral can be expressed in terms of (in)complete elliptic integral of the third kind 
 \begin{equation}
\Pi\left(n= h^2 | m=k^2 \right) = \int_{0}^{1}  \frac{d \chi}{(1-n\,\chi^2)\sqrt{(1-\chi^2)( 1-m\chi^2)}}\,,
\end{equation}
 leading to
 \begin{equation}
\Delta\phi = \frac{2i\left[\left(1+\frac{\zeta}{a}\right)\Pi(n_1| k)-\left(1+\frac{a\,\zeta}{\ell^2}\right)\Pi(n_2|k)\right]}{\sqrt{\chi_{\rm m}^2}\sqrt{1-\frac{b^2}{\ell^2}}}
\end{equation}
with
\begin{equation}
m=k^2= \frac{\chi_{\rm p}^2}{\chi_{\rm m}^2}\,,
\qquad
n_1 = h_1^2 = \chi_{\rm p}^2\,,
\qquad
n_2 = h_2^2 = \frac{a^2}{\ell^2}\chi_{\rm p}^2
\end{equation}
 
 In the case of interest for us $m=k^2< 0$. In order to have periodic motion one should require 
 $$ \Delta\phi = \pi + 2\pi n $$
 this imposes stringent constraints on $r_c$ that are not generically satisfied. As a result, classical motion in the angular variables (for critical radii $r_c$) is non-periodic and as such chaotic / ergodic in the sense that the orbit covers and entire region of the `sphere'. At the quantum level, preventing destructive interference is tantamount to imposing the quantisation condition  $$ \Delta\phi = \pi + 2\pi n $$ that selects only a finite, yet possibly very large, set of  critical radii $r_c$.

 \end{appendix}
\acknowledgments

We thank Alice Aldi, Roberto Benzi, Donato Bini, Davide Bufalini, Davide Cassani, Dario Consoli, Sergio Ferrara, Guido Festuccia, Maurizio Firrotta, Francesco Fucito, Stefano Giusto, Salvo Mancani, Andrei Parnachev, Massimo Porrati, Rodolfo Russo, Congkao Wen for useful discussions and Guido Festuccia, Stefano Giusto, Juan Maldacena, Andrei Parnachev, Rodolfo Russo for their comments on the manuscript.

 
\bibliographystyle{JHEP}
\bibliography{RimChaos}

\providecommand{\href}[2]{#2}\begingroup\raggedright\begin{thebibliography}{10}

\bibitem{Buchdahl:1959zz}
H.~A. Buchdahl, \emph{{General Relativistic Fluid Spheres}},
  \href{http://dx.doi.org/10.1103/PhysRev.116.1027}{\emph{Phys. Rev.} {\bf 116}
  (1959) 1027}.

\bibitem{Cardoso:2008bp}
V.~Cardoso, A.~S. Miranda, E.~Berti, H.~Witek and V.~T. Zanchin,
  \emph{{Geodesic stability, Lyapunov exponents and quasinormal modes}},
  \href{http://dx.doi.org/10.1103/PhysRevD.79.064016}{\emph{Phys. Rev.} {\bf
  D79} (2009) 064016}, [\href{http://arxiv.org/abs/0812.1806}{{\tt
  0812.1806}}].

\bibitem{Maldacena:2015waa}
J.~Maldacena, S.~H. Shenker and D.~Stanford, \emph{{A bound on chaos}},
  \href{http://dx.doi.org/10.1007/JHEP08(2016)106}{\emph{JHEP} {\bf 08} (2016)
  106}, [\href{http://arxiv.org/abs/1503.01409}{{\tt 1503.01409}}].

\bibitem{Shenker:2013pqa}
S.~H. Shenker and D.~Stanford, \emph{{Black holes and the butterfly effect}},
  \href{http://dx.doi.org/10.1007/JHEP03(2014)067}{\emph{JHEP} {\bf 03} (2014)
  067}, [\href{http://arxiv.org/abs/1306.0622}{{\tt 1306.0622}}].

\bibitem{Shenker:2013yza}
S.~H. Shenker and D.~Stanford, \emph{{Multiple Shocks}},
  \href{http://dx.doi.org/10.1007/JHEP12(2014)046}{\emph{JHEP} {\bf 12} (2014)
  046}, [\href{http://arxiv.org/abs/1312.3296}{{\tt 1312.3296}}].

\bibitem{Kiem:1995iy}
Y.~Kiem, H.~L. Verlinde and E.~P. Verlinde, \emph{{Black hole horizons and
  complementarity}},
  \href{http://dx.doi.org/10.1103/PhysRevD.52.7053}{\emph{Phys. Rev.} {\bf D52}
  (1995) 7053--7065}, [\href{http://arxiv.org/abs/hep-th/9502074}{{\tt
  hep-th/9502074}}].

\bibitem{Lunin:2002qf}
O.~Lunin and S.~D. Mathur, \emph{{Statistical interpretation of Bekenstein
  entropy for systems with a stretched horizon}},
  \href{http://dx.doi.org/10.1103/PhysRevLett.88.211303}{\emph{Phys. Rev.
  Lett.} {\bf 88} (2002) 211303},
  [\href{http://arxiv.org/abs/hep-th/0202072}{{\tt hep-th/0202072}}].

\bibitem{Lunin:2001jy}
O.~Lunin and S.~D. Mathur, \emph{{AdS / CFT duality and the black hole
  information paradox}},
  \href{http://dx.doi.org/10.1016/S0550-3213(01)00620-4}{\emph{Nucl. Phys.}
  {\bf B623} (2002) 342--394}, [\href{http://arxiv.org/abs/hep-th/0109154}{{\tt
  hep-th/0109154}}].

\bibitem{Mathur:2005zp}
S.~D. Mathur, \emph{{The Fuzzball proposal for black holes: An Elementary
  review}}, \href{http://dx.doi.org/10.1002/prop.200410203}{\emph{Fortsch.
  Phys.} {\bf 53} (2005) 793--827},
  [\href{http://arxiv.org/abs/hep-th/0502050}{{\tt hep-th/0502050}}].

\bibitem{Mathur:2008nj}
S.~D. Mathur, \emph{{Fuzzballs and the information paradox: A Summary and
  conjectures}},  \href{http://arxiv.org/abs/0810.4525}{{\tt 0810.4525}}.

\bibitem{Hintz:2015jkj}
P.~Hintz and A.~Vasy, \emph{{Analysis of linear waves near the Cauchy horizon
  of cosmological black holes}},
  \href{http://dx.doi.org/10.1063/1.4996575}{\emph{J. Math. Phys.} {\bf 58}
  (2017) 081509}, [\href{http://arxiv.org/abs/1512.08004}{{\tt 1512.08004}}].

\bibitem{Hollands:2019whz}
S.~Hollands, R.~M. Wald and J.~Zahn, \emph{{Quantum Instability of the Cauchy
  Horizon in Reissner-Nordstr\"om-deSitter Spacetime}},
  \href{http://arxiv.org/abs/1912.06047}{{\tt 1912.06047}}.

\bibitem{Bena:2015bea}
I.~Bena, S.~Giusto, R.~Russo, M.~Shigemori and N.~P. Warner, \emph{{Habemus
  Superstratum! A constructive proof of the existence of superstrata}},
  \href{http://dx.doi.org/10.1007/JHEP05(2015)110}{\emph{JHEP} {\bf 05} (2015)
  110}, [\href{http://arxiv.org/abs/1503.01463}{{\tt 1503.01463}}].

\bibitem{Bena:2016agb}
I.~Bena, E.~Martinec, D.~Turton and N.~P. Warner, \emph{{Momentum Fractionation
  on Superstrata}},
  \href{http://dx.doi.org/10.1007/JHEP05(2016)064}{\emph{JHEP} {\bf 05} (2016)
  064}, [\href{http://arxiv.org/abs/1601.05805}{{\tt 1601.05805}}].

\bibitem{Bena:2016ypk}
I.~Bena, S.~Giusto, E.~J. Martinec, R.~Russo, M.~Shigemori, D.~Turton et~al.,
  \emph{{Smooth horizonless geometries deep inside the black-hole regime}},
  \href{http://dx.doi.org/10.1103/PhysRevLett.117.201601}{\emph{Phys. Rev.
  Lett.} {\bf 117} (2016) 201601}, [\href{http://arxiv.org/abs/1607.03908}{{\tt
  1607.03908}}].

\bibitem{Giusto:2009qq}
S.~Giusto, J.~F. Morales and R.~Russo, \emph{{D1D5 microstate geometries from
  string amplitudes}},
  \href{http://dx.doi.org/10.1007/JHEP03(2010)130}{\emph{JHEP} {\bf 03} (2010)
  130}, [\href{http://arxiv.org/abs/0912.2270}{{\tt 0912.2270}}].

\bibitem{Giusto:2011fy}
S.~Giusto, R.~Russo and D.~Turton, \emph{{New D1-D5-P geometries from string
  amplitudes}}, \href{http://dx.doi.org/10.1007/JHEP11(2011)062}{\emph{JHEP}
  {\bf 11} (2011) 062}, [\href{http://arxiv.org/abs/1108.6331}{{\tt
  1108.6331}}].

\bibitem{Bianchi:2016bgx}
M.~Bianchi, J.~F. Morales and L.~Pieri, \emph{{Stringy origin of 4d black hole
  microstates}}, \href{http://dx.doi.org/10.1007/JHEP06(2016)003}{\emph{JHEP}
  {\bf 06} (2016) 003}, [\href{http://arxiv.org/abs/1603.05169}{{\tt
  1603.05169}}].

\bibitem{Bianchi:2017bxl}
M.~Bianchi, J.~F. Morales, L.~Pieri and N.~Zinnato, \emph{{More on microstate
  geometries of 4d black holes}},
  \href{http://dx.doi.org/10.1007/JHEP05(2017)147}{\emph{JHEP} {\bf 05} (2017)
  147}, [\href{http://arxiv.org/abs/1701.05520}{{\tt 1701.05520}}].

\bibitem{Bena:2017upb}
I.~Bena, D.~Turton, R.~Walker and N.~P. Warner, \emph{{Integrability and
  Black-Hole Microstate Geometries}},
  \href{http://dx.doi.org/10.1007/JHEP11(2017)021}{\emph{JHEP} {\bf 11} (2017)
  021}, [\href{http://arxiv.org/abs/1709.01107}{{\tt 1709.01107}}].

\bibitem{Bena:2018mpb}
I.~Bena, E.~J. Martinec, R.~Walker and N.~P. Warner, \emph{{Early Scrambling
  and Capped BTZ Geometries}},
  \href{http://dx.doi.org/10.1007/JHEP04(2019)126}{\emph{JHEP} {\bf 04} (2019)
  126}, [\href{http://arxiv.org/abs/1812.05110}{{\tt 1812.05110}}].

\bibitem{Bena:2019azk}
I.~Bena, P.~Heidmann, R.~Monten and N.~P. Warner, \emph{{Thermal Decay without
  Information Loss in Horizonless Microstate Geometries}},
  \href{http://dx.doi.org/10.21468/SciPostPhys.7.5.063}{\emph{SciPost Phys.}
  {\bf 7} (2019) 063}, [\href{http://arxiv.org/abs/1905.05194}{{\tt
  1905.05194}}].

\bibitem{Bianchi:2018kzy}
M.~Bianchi, D.~Consoli, A.~Grillo and J.~F. Morales, \emph{{The dark side of
  fuzzball geometries}},
  \href{http://dx.doi.org/10.1007/JHEP05(2019)126}{\emph{JHEP} {\bf 05} (2019)
  126}, [\href{http://arxiv.org/abs/1811.02397}{{\tt 1811.02397}}].

\bibitem{Bianchi:2017sds}
M.~Bianchi, D.~Consoli and J.~F. Morales, \emph{{Probing Fuzzballs with
  Particles, Waves and Strings}},
  \href{http://dx.doi.org/10.1007/JHEP06(2018)157}{\emph{JHEP} {\bf 06} (2018)
  157}, [\href{http://arxiv.org/abs/1711.10287}{{\tt 1711.10287}}].

\bibitem{Cardoso:2019rvt}
V.~Cardoso and P.~Pani, \emph{{Testing the nature of dark compact objects: a
  status report}},
  \href{http://dx.doi.org/10.1007/s41114-019-0020-4}{\emph{Living Rev. Rel.}
  {\bf 22} (2019) 4}, [\href{http://arxiv.org/abs/1904.05363}{{\tt
  1904.05363}}].

\bibitem{Barack:2018yly}
L.~Barack et~al., \emph{{Black holes, gravitational waves and fundamental
  physics: a roadmap}},
  \href{http://dx.doi.org/10.1088/1361-6382/ab0587}{\emph{Class. Quant. Grav.}
  {\bf 36} (2019) 143001}, [\href{http://arxiv.org/abs/1806.05195}{{\tt
  1806.05195}}].

\bibitem{Caldarelli:1999xj}
M.~M. Caldarelli, G.~Cognola and D.~Klemm, \emph{{Thermodynamics of
  Kerr-Newman-AdS black holes and conformal field theories}},
  \href{http://dx.doi.org/10.1088/0264-9381/17/2/310}{\emph{Class. Quant.
  Grav.} {\bf 17} (2000) 399--420},
  [\href{http://arxiv.org/abs/hep-th/9908022}{{\tt hep-th/9908022}}].

\bibitem{Chandrasekhar:1985kt}
S.~Chandrasekhar, \emph{{The mathematical theory of black holes}},  in
  \emph{{Oxford, UK: Clarendon (1992) 646 p., OXFORD, UK: CLARENDON (1985) 646
  P.}}, 1985.

\bibitem{Bena:2017xbt}
I.~Bena, S.~Giusto, E.~J. Martinec, R.~Russo, M.~Shigemori, D.~Turton et~al.,
  \emph{{Asymptotically-flat supergravity solutions deep inside the black-hole
  regime}}, \href{http://dx.doi.org/10.1007/JHEP02(2018)014}{\emph{JHEP} {\bf
  02} (2018) 014}, [\href{http://arxiv.org/abs/1711.10474}{{\tt 1711.10474}}].

\bibitem{Kulaxizi:2018dxo}
M.~Kulaxizi, G.~S. Ng and A.~Parnachev, \emph{{Black Holes, Heavy States, Phase
  Shift and Anomalous Dimensions}},
  \href{http://dx.doi.org/10.21468/SciPostPhys.6.6.065}{\emph{SciPost Phys.}
  {\bf 6} (2019) 065}, [\href{http://arxiv.org/abs/1812.03120}{{\tt
  1812.03120}}].

\bibitem{Karlsson:2019qfi}
R.~Karlsson, M.~Kulaxizi, A.~Parnachev and P.~Tadić, \emph{{Black Holes and
  Conformal Regge Bootstrap}},
  \href{http://dx.doi.org/10.1007/JHEP10(2019)046}{\emph{JHEP} {\bf 10} (2019)
  046}, [\href{http://arxiv.org/abs/1904.00060}{{\tt 1904.00060}}].

\bibitem{Festuccia:2005pi}
G.~Festuccia and H.~Liu, \emph{{Excursions beyond the horizon: Black hole
  singularities in Yang-Mills theories. I.}},
  \href{http://dx.doi.org/10.1088/1126-6708/2006/04/044}{\emph{JHEP} {\bf 04}
  (2006) 044}, [\href{http://arxiv.org/abs/hep-th/0506202}{{\tt
  hep-th/0506202}}].

\end{thebibliography}\endgroup
 
\end{document}